\documentclass{aa} %[referee]
\usepackage[varg]{txfonts}
\usepackage{graphicx}

\usepackage{siunitx}
\DeclareSIUnit\erg{erg}
\usepackage{afterpage}

\graphicspath{ {figures/} }

\begin{document}
\title{Studying the ISM at $\sim$ 10 pc scale in NGC 7793 with MUSE - I. Data description and properties of the ionised gas}
\author{Lorenza Della Bruna\inst{\ref{inst1}}
\and Angela Adamo\inst{\ref{inst1}}
\and Arjan Bik\inst{\ref{inst1}} 
\and Michele Fumagalli\inst{\ref{inst2}, \ref{inst3}, \ref{inst4}} 
\and Rene Walterbos\inst{\ref{inst5}}  
\and Göran Östlin\inst{\ref{inst1}}  
\and Gustavo Bruzual\inst{\ref{inst6}}
\and Daniela Calzetti\inst{\ref{inst7}}
\and Stephane Charlot\inst{\ref{inst8}} 
\and Kathryn Grasha\inst{\ref{inst9}}
\and Linda J. Smith\inst{\ref{inst10}} 
\and David Thilker\inst{\ref{inst11}}  
\and Aida Wofford\inst{\ref{inst12}}  
}
%\and Lauren Bittle\inst{\ref{inst2}}} 

\institute{Department of Astronomy, Oskar Klein Centre, Stockholm University, AlbaNova University
Centre, SE-106 91 Stockholm, Sweden\label{inst1}
\and
Institute for Computational Cosmology, Durham University, South Road, Durham, DH1 3LE, UK\label{inst2}
\and
Centre for Extragalactic Astronomy, Durham University, South Road, Durham, DH1 3LE, UK\label{inst3}
\and
Dipartimento di Fisica G. Occhialini, Universit\`a degli Studi di Milano Bicocca, Piazza della Scienza 3, 20126 Milano, Italy\label{inst4}
\and
Department of Astronomy, New Mexico State University, Las Cruces, NM, 88001, USA\label{inst5}
\and
Instituto de Radioastronomía y Astrofísica, UNAM, Campus Morelia, Michoacan, C.P. 58089, México \label{inst6}
\and
Department of Astronomy, University of Massachusetts, Amherst, MA 01003, USA\label{inst7}
\and
Sorbonne Université, CNRS, UMR7095, Institut d'Astrophysique de Paris, F-75014, Paris, France\label{inst8}
\and
Research School of Astronomy and Astrophysics, Australian National University, Canberra, ACT 2611, Australia\label{inst9}
\and
Space Telescope Science Institute and European Space Agency, 3700 San Martin Drive, Baltimore, MD 2121, USA\label{inst10}
\and
Department of Physics and Astronomy, Johns Hopkins University, 3400 N. Charles Street, Baltimore, MD 21218, USA\label{inst11}
\and
Instituto de Astronom\'ia, Universidad Nacional Aut\'onoma de M\'exico, Unidad Acad\'emica en Ensenada, Km 103 Carr. Tijuana$-$Ensenada, Ensenada 22860, M\'exico\label{inst12}
}

\authorrunning{Della Bruna et al.}
\titlerunning{Studying the ISM at $\sim$ 10 pc scale in NGC 7793}

\date{Received XXX / Accepted YYY}

\abstract
% Context
{Studies of nearby galaxies reveal that around 50\% of the total H$\alpha$ luminosity in late-type spirals originates from diffuse ionised gas (DIG), which is a warm, diffuse component of the interstellar medium that can be associated with various mechanisms, the most important ones being 'leaking’ HII regions, evolved field stars, and shocks.}
% Aim
{Using MUSE Wide Field Mode adaptive optics-assisted data, we study the condition of the ionised medium in the nearby (D = 3.4 Mpc) flocculent spiral galaxy NGC 7793 at a spatial resolution of $\sim$ 10 pc. We construct a sample of HII regions and investigate the properties and origin of the DIG component.}
% Methods
{We obtained stellar and gas kinematics by modelling the stellar continuum and fitting the H$\alpha$ emission line. We identified the boundaries of resolved HII regions based on their H$\alpha$ surface brightness. As a way of comparison, we also selected regions according to the H$\alpha$/[SII] line ratio; this results in more conservative boundaries. Using characteristic line ratios and the gas velocity dispersion, we excluded potential contaminants, such as supernova remnants (SNRs) and planetary nebulae (PNe). The continuum subtracted HeII map was used to spectroscopically identify Wolf Rayet stars (WR) in our field of view. Finally, we computed electron densities and temperatures using the line ratio [SII]6716/6731 and [SIII]6312/9069, respectively. We studied the properties of the ionised gas through ‘BPT’ emission line diagrams combined with velocity dispersion of the gas.}
%Results
{We spectroscopically confirm two previously detected WR and SNR candidates and report the discovery of the other seven WR candidates, one SNR, and two PNe within our field of view. The resulting DIG fraction is between  $\sim$ 27 and 42\% depending on the method used to define the boundaries of the HII regions (flux brightness cut in H$\alpha$ = $\SI{6.7e-18}{\erg \per \second \per \centi \metre \squared}$ or H$\alpha$/[SII] = 2.1, respectively). In agreement with previous studies, we find that the DIG exhibits enhanced [SII]/H$\alpha$ and [NII]/H$\alpha$ ratios and a median temperature that is $\sim$ 3000 K higher than in HII regions. We also observe an apparent inverse correlation between temperature and H$\alpha$ surface brightness. In the majority of our field of view, the observed [SII]6716/6731 ratio is consistent within 1$\sigma$ with $n_e$ < 30 cm$^{-3}$, with an almost identical distribution for the DIG and HII regions.
The velocity dispersion of the ionised gas indicates that the DIG has a higher degree of turbulence than the HII regions. Comparison with photoionisation and shock models reveals that, overall, the diffuse component can only partially be explained via shocks and that it is most likely consistent with photons leaking from density bounded HII regions or with radiation from evolved field stars.
Further investigation will be conducted in a follow-up paper.}
{}
\maketitle

\keywords{Galaxies: individual: NGC 7793 - Galaxies: ISM - ISM: structure -  HII regions}

\section{Introduction}
\label{section:introduction}
% Stellar feedback
Stellar feedback is an essential mechanism for the regulation of star formation. This effect, which is observed at galactic scales, originates at much smaller scales, around young, massive stars, that emit powerful ionising radiation and inject mechanical energy into the interstellar medium (ISM), thus heating it and therefore preventing catastrophic star formation~\citep{cole00}.
The link between large (kpc) and small (10s of parsec) scales is not yet well understood. With the recent development of integral field spectroscopy, nowadays, it is possible to resolve ionised gas in nearby galaxies and gather, at the same time, information about its properties, ionisation state, and kinematics.

% DIG observation
\par In the past two decades, it has been established that around 50\% (with, however, a large scatter) of the total H$\alpha$ luminosity in late-type spiral galaxies originates from a warm ($T \sim 6000 - \SI{10000}{\kelvin}$), diffuse ($n_e \sim 0.03 - \SI{0.08}{\per \cubic \centi \metre}$) component of the ISM \citep{ferguson96, hoopes96, zurita00, thilker02, oey07}. This diffuse ionised gas (DIG; also frequently referred to also as a warm ionised medium, WIM) has since been the intense object of study, both in the Milky Way and in nearby galaxies \citep[for a review see][]{mathis00, haffner09}. Two of the main observational pieces of evidence are that: first, the DIG has a lower ionisation stage with respect to HII regions. In particular, it has been observed that ratios of the forbidden lines [SII]6716 and [NII]6584 with respect to H$\alpha$ are significantly larger in the DIG than in classical HII regions \citep{madsen06}.
Moreover, in edge-on galaxies these ratios are observed to increase with increasing height above the midplane~\citep{rand98}. The second piece of evidence is that the DIG temperature is higher than in classical HII regions \citep[$\Delta$ T = 2000 K on average;][]{haffner99,madsen06}.

% DIG - origin?
The origin of this diffuse component is still not fully understood. The main candidates are leaking HII regions~\citep{zurita02, weilbacher18},
evolved field stars \citep{hoopes00,zhang17},
shocks~\citep{collins01},
and cosmic rays~\citep{vandenbroucke18}.
\citet{zurita00} argued that the DIG distribution very often spatially coincides with the location of the HII regions and that the correlation is stronger for the most luminous regions. In a follow-up work, \citet{zurita02} showed that a pure density bounded HII region scenario allows for a reasonable prediction of the spatial distribution of the DIG. If the DIG is mainly or solely ionised by radiation leaking from HII regions, the observed enhanced line ratios [SII]6716/H$\alpha$ and [NII]6584/H$\alpha$, indicating a lower ionisation stage for these elements, still have yet to be explained. Ionising photons can escape either via empty holes in the neutral gas surrounding an HII region or from a region that is density bounded. In the second case, the resulting spectrum is modified because of partial absorption of the radiation as it travels through neutral H and He. Photoionisation models from \citet{hoopes03} and \citet{wood04} showed that processing of the radiation by a stratified interstellar medium can result in a spectrum that becomes increasingly harder with distance from the source between the HI and HeI ionisation edge, and softer at higher energies. This naturally results in an increasing temperature with distance from the source of ionisation, and it explains the observed increase in [SII]6716/H$\alpha$ and [NII]6584/H$\alpha$ above the midplane. Nevertheless, these models require  some additional non-ionising heating in order to match the largest line ratios observed.

\par Of particular interest for the study of nearby star forming regions are spectrographs equipped with integral field units (IFUs) and imaging fourier transform spectrographs (IFTS), such as the Multi Unit Spectroscopic Explorer (MUSE) at VLT~\citep{bacon10} and the SITELLE instrument at CFHT~\citep{drissen19}. There are several ongoing surveys of nearby galaxies using these instruments: of particular relevance for the study of stellar feedback at small scales are the MUSE Atlas of Discs (MAD) survey~\citep{erroz-ferrer19, denbrok19}, probing scales of $\sim$ 100 pc across a sample of 45 disc galaxies at an average distance of 20 Mpc, and the Physics at High Angular resolution in Nearby GalaxieS (PHANGS) survey (Leroy et al. in prep.), which is targeting 19 galaxies with MUSE at distance d $\sim 14$ Mpc, reaching scales of $\sim 40$ pc. At even smaller scales, the Star formation, Ionised Gas, and Nebular Abundances Legacy Survey \citep[SIGNALS;][]{rousseau19} will observe over 50000 resolved HII regions in nearby (< 10 Mpc) galaxies, with a spatial resolution ranging from 2 to 40 pc with SITELLE. These surveys will allow for a statistical study of the properties of HII regions and of the ionised gas surrounding them.

\par Our aim with this work and future publications, is to target a few local galaxies (d $\sim$ 5 Mpc) for which a detailed census of the stellar populations (with HST) and maps of the molecular gas (with ALMA) are accessible at a very high spatial resolution. Combining the information provided by MUSE on the condition of the ionised ISM with these multiwavelength datasets, enables us to study the entire star formation cycle from parsec to galactic scales and from the very early phases of star formation to the creation of HII-regions and a warm ionised medium.

%%%%%% Aim of this work
\par We present recent MUSE AO observations of the nearby spiral galaxy NGC 7793. We use the extensive spectral information provided by MUSE to study the ionisation state of the gas: we identify the boundaries of HII regions and then investigate the properties of the DIG, in an attempt to shed some light on the origin of this diffuse component and on the process of stellar feedback.  In a followup paper (hereafter Paper II; Della Bruna et al in prep.), we will combine the MUSE results on the ionised gas with the stellar and cluster population studied within the HST--LEGUS programme and the dense gas properties provided by the ALMA-LEGUS data, to provide a full picture of the star formation cycle at scales of $\sim$ 10 pc.

%%%%%% Structure of the paper
\par This paper is organised as follows: in Sect.~\ref{section:data} we introduce the data and in Sect.~\ref{section:data_reduction} we summarise the reduction process. In Sects.~\ref{section:kinematics} and~\ref{section:reddening} we present maps of the stellar and gas kinematics and of interstellar extinction.
In Sect.~\ref{section:analysis_gas} we identify the boundaries of HII regions and compile a catalogue of candidate planetary nebulae, supernova remnants and Wolf Rayet stars.
In Sect.~\ref{section:ne_Te} we compute the electron temperature and density and in Sect.~\ref{section:dig_properties} we compare HII regions and DIG regions with photoionisation models and model grids for strong shocks. Finally, in Sect.~\ref{section:conclusions} we draw our conclusions.

\section{Data description}
\label{section:data}
\begin{figure*}[!t]
%\center
\includegraphics[width=18cm]{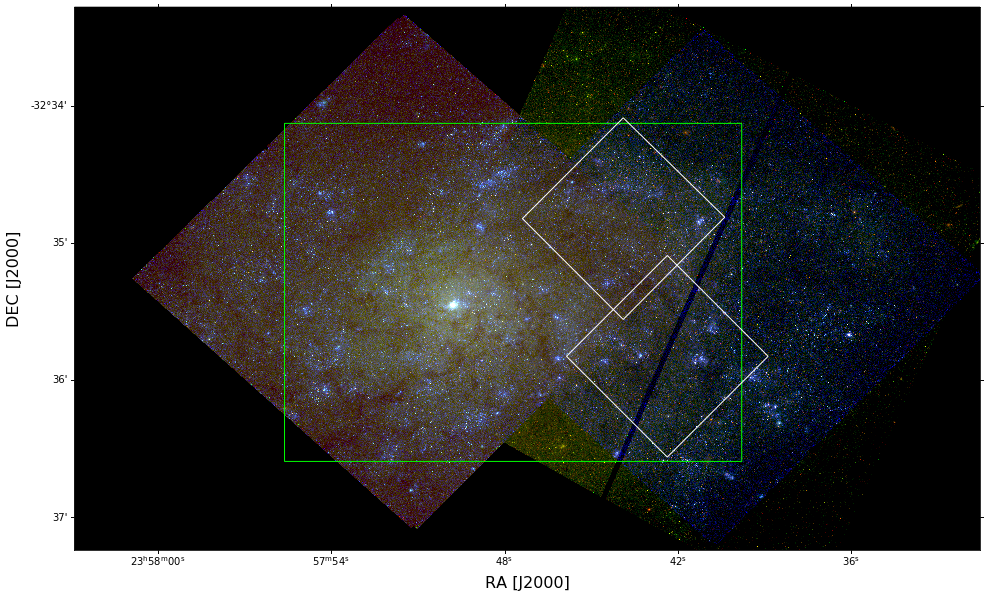} \\
\includegraphics[width=9cm]{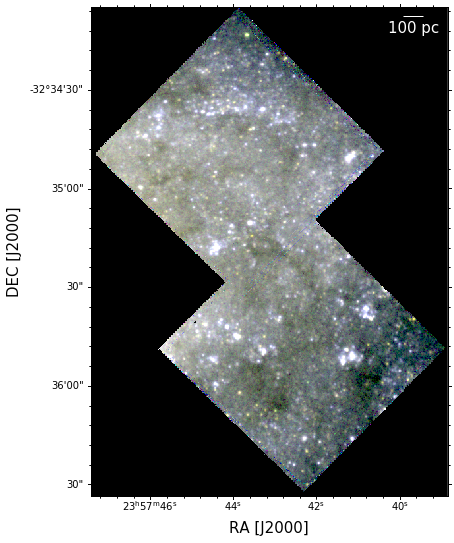}
\includegraphics[width=9cm]{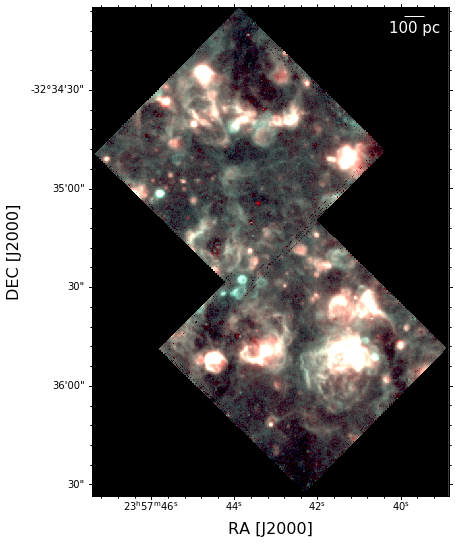}
\caption{Upper panel: Three-colour image from the HST LEGUS data (R: F814W, G: F555W, B: F336W). The white contours indicate the location of the two MUSE pointings, and the green rectangle shows the coverage of the ALMA CO(2-1) data. Bottom: RGB composite of stellar continuum bands (left panel; R: 6750 - 6810~\AA{}, G: 6520 - 6528~\AA{}, B: 
4875 - 4950~\AA{}) and of gas emission lines (right panel; R: H$\alpha$, G: [SII]6716,6731, B: [OIII]4959,5007) obtained from the two MUSE tiles.}
\label{fig:pointings}
\end{figure*}

Our target is NGC 7793, a flocculent spiral galaxy (SAd) at a distance $\simeq$ 3.4 Mpc \citep{zgirski17}, located in the Sculptor group. Its main properties are summarised in Table~\ref{table:ngc7793_param}. NGC 7793 is part of the sample of galaxies studied by the Hubble Space Telescope (HST) Treasury programme LEGUS\footnote{HST GO–13364.}~\citep{calzetti15}; the dataset is described in~\citet{grasha18}. Recent ALMA data in the CO (J = 2–1) transition provide us with information about the giant molecular cloud (GMC) population at a $0.85''$ resolution, and detection limits corresponding to a minimum cloud mass of $10^4 M_{\sun}$~\citep{grasha18}. A three-colour image from HST and the coverage of the ALMA dataset are shown in Fig.~\ref{fig:pointings}.

Within the field of view (FoV) of these two datasets, we have obtained IFU spectroscopy with the MUSE instrument at the ESO Very Large Telescope (VLT) observatory, as part of the Adaptive Optics Science Verification run\footnote{ESO programme 60.A-9188(A), PI Adamo.}. The data were acquired in the wide field mode (WFM) configuration with the extended wavelength setting, covering the spatial extent of 1 arcmin$^2$ with $0.2 \arcsec$ sampling and the spectral range 4650 - 9300~\AA{} at 1.25~\AA{} sampling. In the extended wavelength AO mode, the wavelength range $\sim$ 5760 - 6010~\AA{} is blocked in order to avoid contamination by sodium light from the laser guide system; at our redshift, this rules out several important lines such as the [NII]5755 `auroral' line used for temperature diagnostics. The spectral resolution of the instrument $R = \lambda/\Delta \lambda$ ranges from R = 1770 in the blue end to R = 3590 in the red end of the spectrum. On the final data product, we measure a seeing of $0.71 \arcsec$ FWHM (at 4900~\AA{}). 

\par The observation consists of two pointings (centred at R.A. 23:57:42.307, Dec. -32:35:48.150 and R.A. 23:57:43.889, Dec. -32:34:49.353, see Fig.~\ref{fig:pointings}), each observed during four exposures, for a total of $2100~\rm s$ on source.
In order to take advantage of the seeing enhancement AO performance of MUSE, a Tip Tilt `stellar' like source with brightness in $R < 17.5$ mag is required within 105" of the MUSE FoV and not closer than 52". The only bright source that could satisfy these technical constraints and allow us to fit the MUSE FoV within the HST--LEGUS coverage and ALMA--LEGUS coverage was the nuclear star cluster in the centre of the galaxy. Four sky frames of $120$ s exposure were also acquired in an object-sky-object pattern, located at a distance $\sim$ 3.8 kpc south-west of the science pointings.

\begin{table}
\caption{Properties of NGC 7793.}
\begin{tabular}{lcc}
\hline \hline
Parameter & Value & Ref. \\ \hline
Morphological Type & SA(s)d & (1) \\
Systemic velocity & 227 km/s & (2) \\
Inclination angle & 47$^\circ$ & (2) \\
Stellar mass & \num{3.2e9} $M_\sun$ & (2) \\
SFR(UV) & 0.52 $M_\sun$/yr & (3)\\
Central 12 + log(O/H) & 8.50 $\pm$ 0.02 & (4)  \\
12 + log(O/H) gradient & -0.0662 $\pm$ -0.0104 dex/kpc & (4) \\ 
\hline
\end{tabular}
\tablebib{(1) \citet{devaucouleurs91}; (2) \citet{carignan90}; (3) \citet{calzetti15}; (4) \citet{pilyugin14} }
\label{table:ngc7793_param}
\end{table}

\section{Data reduction}
\label{section:data_reduction}

\subsection{Instrumental signature, flux calibration, and data alignment}
\label{section:muse_pipeline}
The data have been reduced using the MUSE ESO pipeline \citep[v2.2]{weilbacher14}. We follow the first conventional reduction steps: we combine flat and bias frames into a master frame, we perform wavelength calibration and compute the line spread function (LSF) from the arc-lamps frames. We use a pre-computed geometry table to determine the location of the IFU slices on the FoV and we perform illumination correction combining twilight flats. Using the pipeline recipe \texttt{muse\_scibasic}, we then combine all this information to remove the instrument signature from science exposures, sky frames and standard star. We perform flux calibration using the standard star and astrometry calibration using the astrometric catalogue distributed with the pipeline. We finally run a first time \texttt{muse\_scipost} on the object frames to apply these calibrations and remove cosmic ray signatures.
\par As a last step, we create the astrometry for the final datacube. Since we are performing sky subtraction with external software (see Sect. \ref{section:sky_subtraction}) operating on datacubes rather than on a `pipeline-friendly', pixeltable level, extra steps need to be taken. We match the absolute astrometry of each pointing to the HST LEGUS observations \citep{grasha18}, use the pipeline recipe \texttt{muse\_exp\_align} to compute the relative shifts of the various exposures and produce a cube having the correct astrometry using \texttt{muse\_exp\_combine}. We then take a step back and manually adjust the headers of the non-reduced object pixel tables from \texttt{muse\_scibasic} in order to correct for the spatial shift. Finally, we run a second time \texttt{muse\_scipost}, using the WCS information from the combined cube to ensure that each of the exposures is added on the right location on the output frame. In this way, once the single science exposures are sky-corrected, the corresponding datacubes can be merged together by a simple exposure time-weighted average.

\subsection{Improved sky subtraction}
\label{section:sky_subtraction}
We perform sky subtraction using \textsc{zap}~\citep[v2.0]{soto16}. This software implements a principal component analysis (PCA) on each sky frame, that is it decomposes the image in a set of orthogonal vectors, each accounting for the largest possible variance in the data. It then applies this vector basis to the science image, finds the corresponding eigenvalues for each individual spectrum and finally uses this to subtract the sky signature from the science frame.

\par We reduce the four sky exposures with similar steps as for the science frames using the \texttt{muse\_scipost} pipeline recipe, and obtain four sky cubes. We then create a skymask, to remove eventual contamination from galactic HII regions and bad pixels. We furthermore adopt two additional precautions when running \textsc{zap}. In the first place, we exclude the bluest end of the spectrum (4600 - 4812~\AA{}) when computing the eigenvectors on the sky cube, as the LSF varies strongly in this range. The excluded wavelength range was determined by inspecting the behaviour of the first 40 principal vectors; for this portion of the spectrum, we compute sky subtraction using the standard pipeline recipe. Secondly, since the software does not perform accurately in the presence of strong emission lines, we clip the most prominent gas emission lines from the target galaxy (H$\alpha$, H$\beta$, [OIII]4959,5007, [NII]6548, 6584) before computing the eigenvalues of the sky vectors on the data.
We then run \textsc{zap} on the four sky frames and apply to each science exposure the correction corresponding to the nearest sky frame in time. We carefully inspect the resulting principal vectors and variance ratios, as well as the subtracted sky emission to ensure no science is subtracted. Finally, we combine the so obtained eight sky subtracted frames using an exposure time weighted average. By comparing the resulting sky subtracted datacube to the one obtained with the standard MUSE pipeline sky subtraction, we observe that our own sky subtraction significantly improves the quality of the final cubes with residual sky lines strongly reduced. 

\section{Stellar and gas kinematics}
\label{section:kinematics}

\subsection{Stellar continuum modelling}
\label{section:spectral fitting}
In order to correct for stellar absorption, in particular below the Balmer lines, we perform a fit to the stellar continuum. This also allows us to obtain information on the stellar kinematics (see Sect.~\ref{section:v_sigma}). To achieve a sufficiently high signal to noise in the continuum, we spatially bin the data using the Voronoi tessellation method described in~\citet{cappellari03}. We use the weighted Voronoi tesselation (WVT) adaptation of the algorithm proposed by~\citet{diehl06}. We create several patterns with a signal-to-noise-ratio (S/N) $\sim$ 40 - 200 in the continuum region 5020 - 5060~\AA{} and a maximum cell width of 100 pixels. We find that the best compromise between goodness of fit and number of bins is achieved with a S/N = 160 over the continuum region (corresponding to a S/N $\sim$ 40 in narrow stellar absorption features). We then perform spectral fitting using \textsc{pPXF}~\citep{cappellari04,cappellari17}.

This software models a galaxy spectrum as
$$G_{mod} (x) = T(x) * \mathcal{L} (c x),$$
where $x = \ln \lambda$, $T$ are template spectra and $\mathcal{L}$ is the line of sight velocity distribution (LOSVD) of the stars. The underlying assumption is that any difference between the galaxy spectrum and the templates is purely due to broadening by the LOSVD. The template resolution is matched to the instrumental one by convolving the templates with a kernel $\kappa$ defined by
$$LSF_{instrument} (x, \lambda) = LSF_{templates} (x, \lambda) * \kappa (x, \lambda).$$
The best fitting template is determined by maxmizing a penalised likelihood function, with a penalty function describing the deviation of the LOSVD from a pure Gaussian shape. We assume a Gaussian instrumental LSF\footnote{We use the parametrization of the MUSE LSF by~\citet{guerou17}:
$\sigma_{LSF} = \num{6.266e-8} \lambda^2 - \num{9.824e-4} \lambda + 6.286.$} and use the most recent simple stellar populations from Charlot \& Bruzual (in preparation). In the range 3540 - 7410~\AA{}, these models are largely based on the MILES stellar library~\citep{blazquez06} with FWHM $\sim$ 2.5~\AA{} spectral resolution~\citep{barroso11} and are extended using synthetic stellar spectra to cover the spectral range from 5.6~\AA{} to 3.6 cm as well as hot stars not available in the MILES library~\citep[see][]{vidalgarcia17,plat19,werle19}. These models follow the PARSEC stellar evolutionary tracks~\citep{bressan12, chen15}, which include the evolution of both Wolf Rayet and TP-AGB stars, to describe the evolution of stellar populations of 16 metallicities ranging from Z = 0.0005 to 3.5 $Z_\sun$ in the age range from 1 Myr to 14 Gyr in 220 steps. The advantage of these SSP models over other models based on the MILES library is the inclusion of stellar populations younger than 60 Myr,
which are crucial to study the young star-forming regions in our FoV.

\par We run \textsc{pPXF} on the Voronoi tessellated datacube, masking all the relevant emission lines in the MUSE range, as well as sky subtraction residuals. However, as the red end of the data is too heavily affected by residual sky emission, we limit our fit to the wavelength range 4600 - 7300~\AA{}. We obtain a best-fit to the continuum in each Voronoi bin; in order to determine the stellar population at a spaxel resolution, we then make the simplifying assumption that the stellar population spectrum does not vary - up to a scale factor - within each bin. We find the best scale factor by computing the median distance between the spectrum in each spaxel and the one in the corresponding Voronoi bin (masking all gas emission lines). We then subtract the contribution from the stars and obtain a datacube consisting exclusively of gas emission (in the following we refer to this data product as a `gas cube'). A three-colour composite image obtained from the gas cube is shown in Fig.~\ref{fig:pointings}.

\subsection{Emission lines fitting}
\label{section:el_fitting}
We independently fit gas emission lines, using a single component Gaussian resulting from a convolution of the instrumental LSF with a Gaussian intrinsic broadening profile. We observe that the reduced $\chi^2$ goodness-of-fit is uniform over the whole FoV (with the exception of the brightest HII regions peaks, where the S/N is exceptionally large), indicating that a single component fit is a valid approximation for both the DIG and HII regions within the data. Analogously to the stellar continuum fit (see Sect.~\ref{section:spectral fitting}), we Voronoi tessellate the datacube to increase the S/N of the lines.
From the gas cube, we produce linemaps by integrating the flux in the spectral region around each line, and use these to create several Voronoi tessellation patterns that we then apply to the cube. We select a tessellation pattern based on our needs: whenever we are performing an analysis for which a reddening correction is required, we create a pattern corresponding to a S/N in H$\beta \sim 20$, a threshold determined by requiring a sufficiently small uncertainty in the extinction determined from the Balmer decrement (see Sect.~\ref{section:reddening}); otherwise, we require a S/N $\sim 20$ in the weakest line of interest. The selected pattern is applied to the datacube before fitting the emission lines required for the analysis.

\subsection{Velocity and velocity dispersion maps}
\label{section:v_sigma}

\begin{figure*}
%\center
\includegraphics[width=9cm]{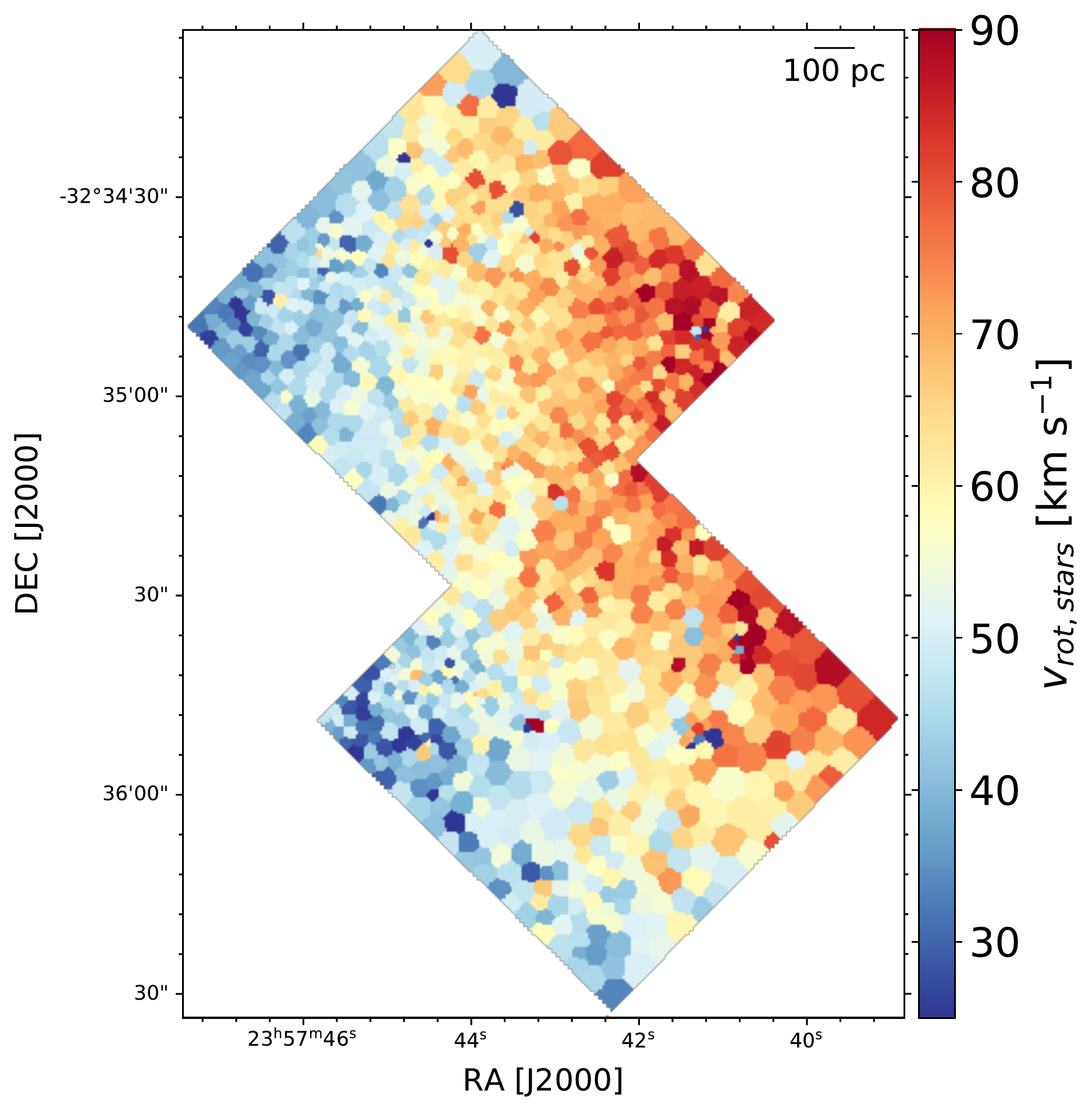}
\includegraphics[width=9cm]{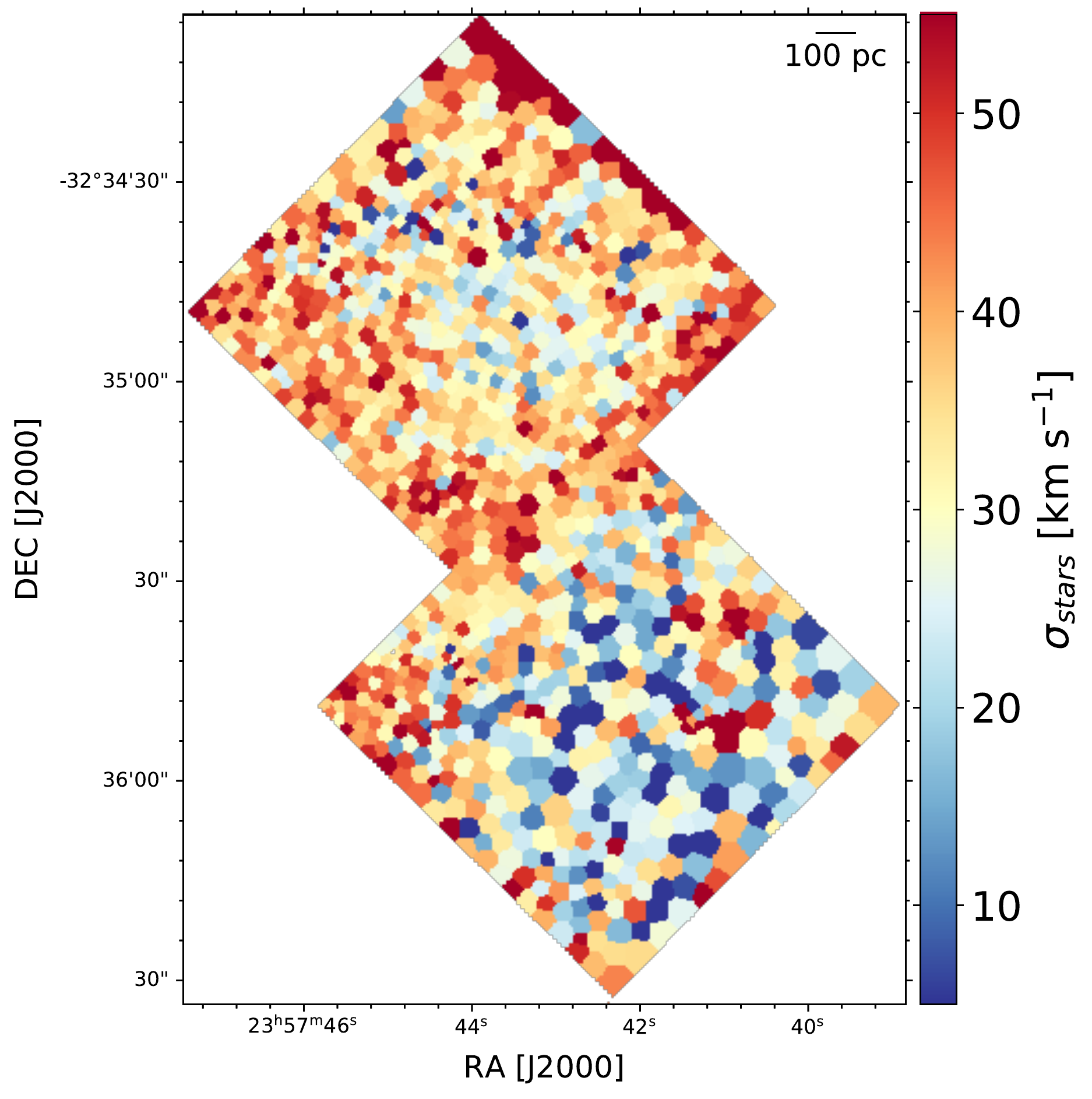}
\caption{Stellar $v$ and $\sigma$ parameters of the LOSVD recovered by fitting the stellar continuum with pPXF.}
\label{fig:stellar_kinematics}
\end{figure*}

\begin{figure*}
%\center
\includegraphics[width=9cm]{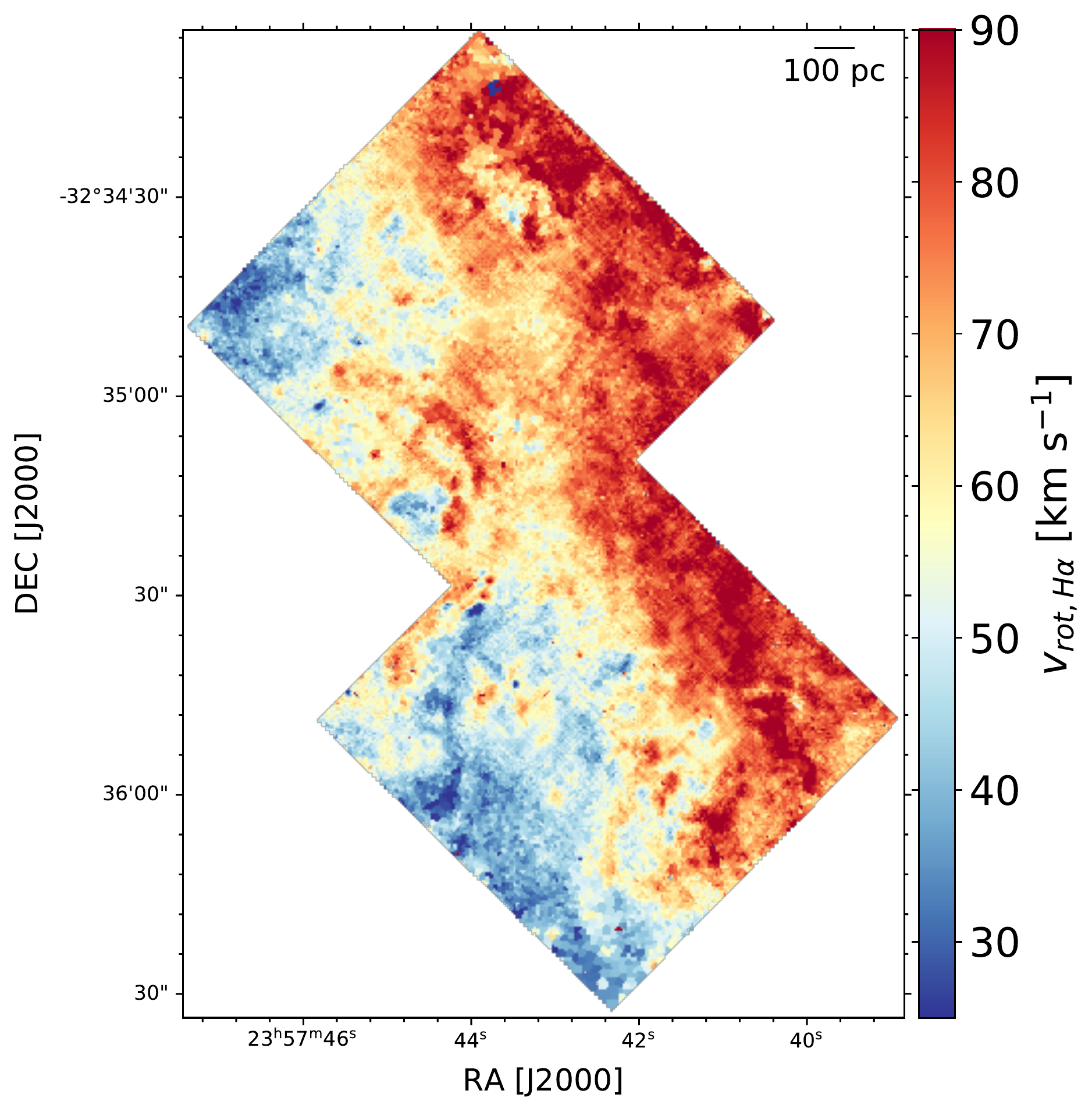}
\includegraphics[width=9.2cm]{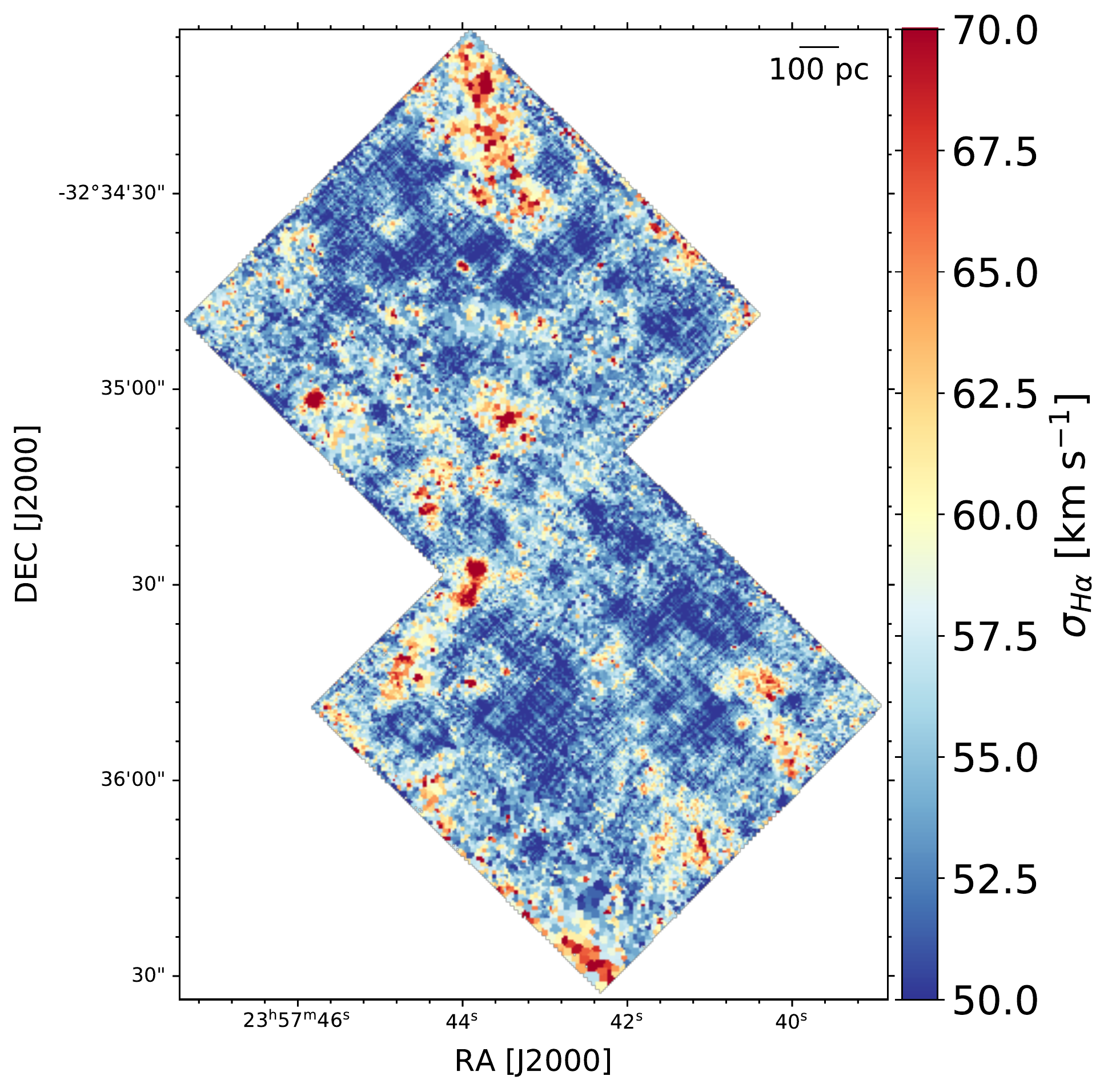}
\caption{Velocity and velocity dispersion of the H$\alpha$ line obtained from a single Gaussian component fit to the line. The instrumental LSF sets a limit for the lowest gas velocity dispersion measurable $\sim 50$ km/s.}
\label{fig:halpha_kinematics}
\end{figure*}

In~Fig.~\ref{fig:stellar_kinematics} we show the stellar LOSVD parameters $v, \sigma$ resulting from the fit to the stellar continuum (see Sect. \ref{section:spectral fitting}). Fig.~\ref{fig:halpha_kinematics} shows the morphology and kinematics of the H$\alpha$ line, obtained by fitting a single Gaussian component to the line (Sect. \ref{section:el_fitting}). With the assumed instrumental LSF parametrization (see Sect.~\ref{section:spectral fitting}), LSF($\lambda_{H\alpha,obs}) \approx 2.54$ \AA{}. Since emission lines have a total broadening
$$\sigma_{tot} = \sqrt{\sigma_{LSF}^2 + \sigma_{gas}^2},$$
this sets a lower limit to the gas velocity dispersion we can measure to $\sim$ 1~\AA{} $\equiv$ 46~km/s. The observed velocity is corrected for an inclination $i$ of 47 degrees and a systemic velocity of 227 km/s \citep{carignan90} as follows:
$$v_{rot} = (v_{obs} - v_{syst}) / sin(i).$$ The difference in spatial resolution (caused by a different Voronoi tessellation) between stars (Fig.~\ref{fig:stellar_kinematics}) and gas (Fig.~\ref{fig:halpha_kinematics}) is due to the necessity of reaching an optimal S/N on the continuum and emission line respectively before fitting.

The velocity fields of stars and gas (left panels of Fig.~\ref{fig:stellar_kinematics} and \ref{fig:halpha_kinematics} respectively) show similar features; we are observing the receding side of the galaxy, and we can see the rotational velocity increase towards the outer part of the disc. The observed velocity range is in agreement with the previous deep H$\alpha$ observations by~\citet{dicaire08}. The stellar velocity dispersion (right panel of Fig.~\ref{fig:stellar_kinematics}) does not show sharp variations over the whole FoV, but is generally lower in the southern pointing, probably due to the presence of the largest HII region complexes, affecting the stellar light. In the velocity dispersion map of the H$\alpha$ line we observe instead some regions with a higher $\sigma$, clearly standing out against the background, and not coincident with HII regions; we investigate their origin in Sect.~\ref{section:outliers}.

\section{Reddening estimate}
\label{section:reddening}

We estimate a reddening map using \textsc{Pyneb}~\citep{Luridiana15} and the H$\alpha$/H$\beta$ ratio, obtained from fitting the gas cube tessellated to a S/N $\sim$ 20 in H$\beta$. We assume an extinction law $k(\lambda)$ from~\citet{cardelli89} and a theoretical Balmer decrement $(H\alpha/H\beta)_{int} = 2.863$, corresponding to case B recombination with $T_e = \SI{1e4}{\kelvin}$ and $n_e = \SI{1e2}{\centi \metre}^{-3}$~\citep{osterbrock}.
It has to be remarked that this assumption describes a foreground screen for the dust, adequate for HII region emission but less accurate in case of spatially extended emission, as it can be the case in the DIG. However, we would like to point out that in this work a reddening estimate is used only to correct the [OIII]/H$\alpha$ line ratio when identifying candidate PNe.

\par
\begin{figure}
\center
\includegraphics[width=9cm]{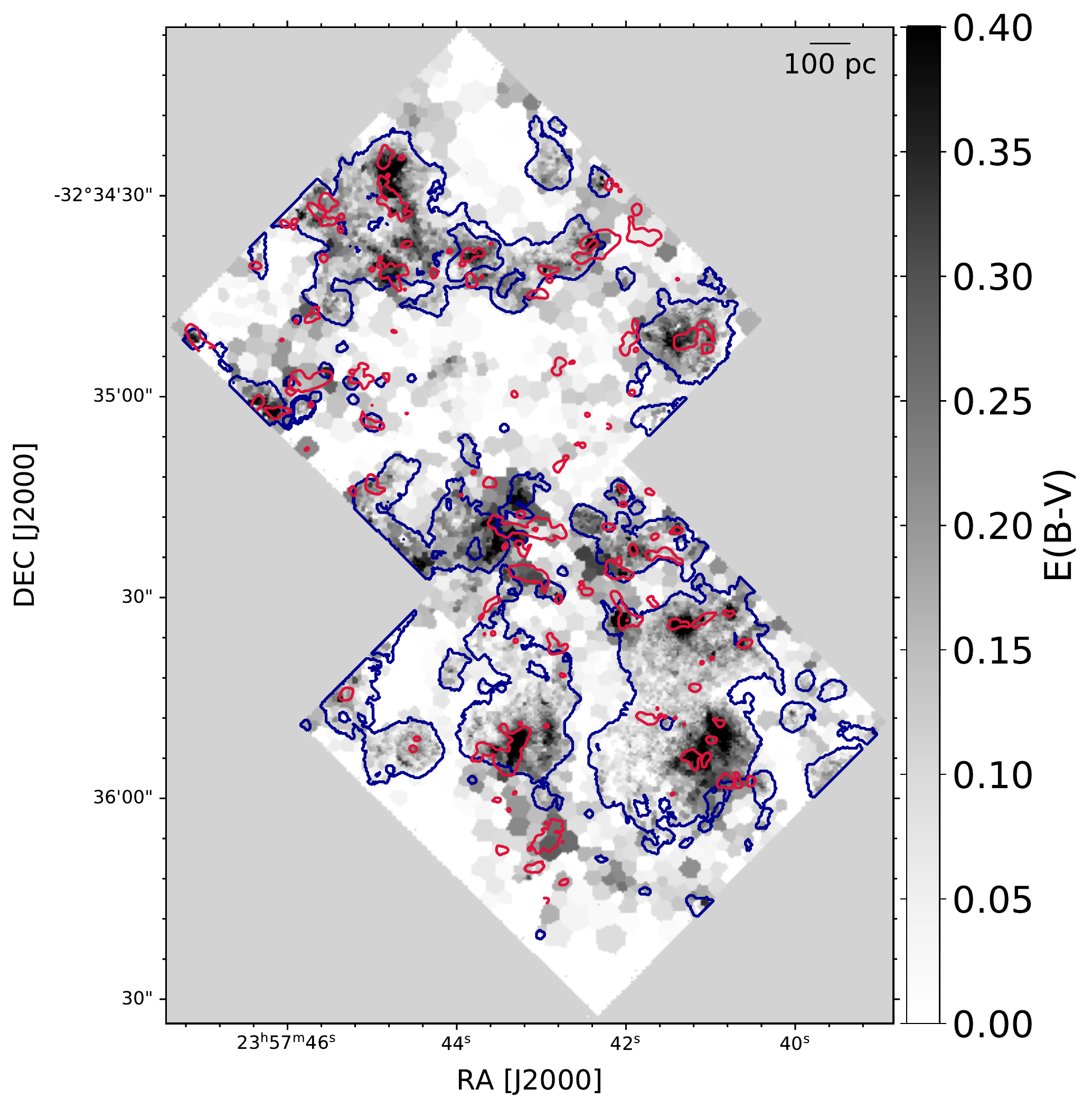}
\caption{E(B-V) extinction map. Overlaid are the contours of GMCs detected with ALMA (red) and of HII regions (blue; this work).}
\label{fig:extinction_map}
\end{figure}

 In Fig.~\ref{fig:extinction_map} we show the resulting extinction map E(B-V), with overlaid GMC contours from the ALMA CO(2-1) data. We can observe that the position of the GMCs is clearly correlated with the higher extinction measured from the Balmer decrement. By comparing this map with the outer boundaries (blue contours) of HII regions (determined in Sect.~\ref{section:hiiregions_selection}), we observe that dense gas clouds with a mass above $10^4 M_{\sun}$ are still coincident (along the line of sight) with many of the HII regions present in the FoV, suggesting that the dense and ionised gas phases can coexist during the first Myr of evolution. Some regions exhibit a strong extinction but no star formation: these are the locations where the next generation of stars is expected to form.

\section{Structure of the ionised gas}
\label{section:analysis_gas}

\subsection{HII regions selection}
\label{section:hiiregions_selection}
\begin{figure}
\center
\includegraphics[width=9cm]{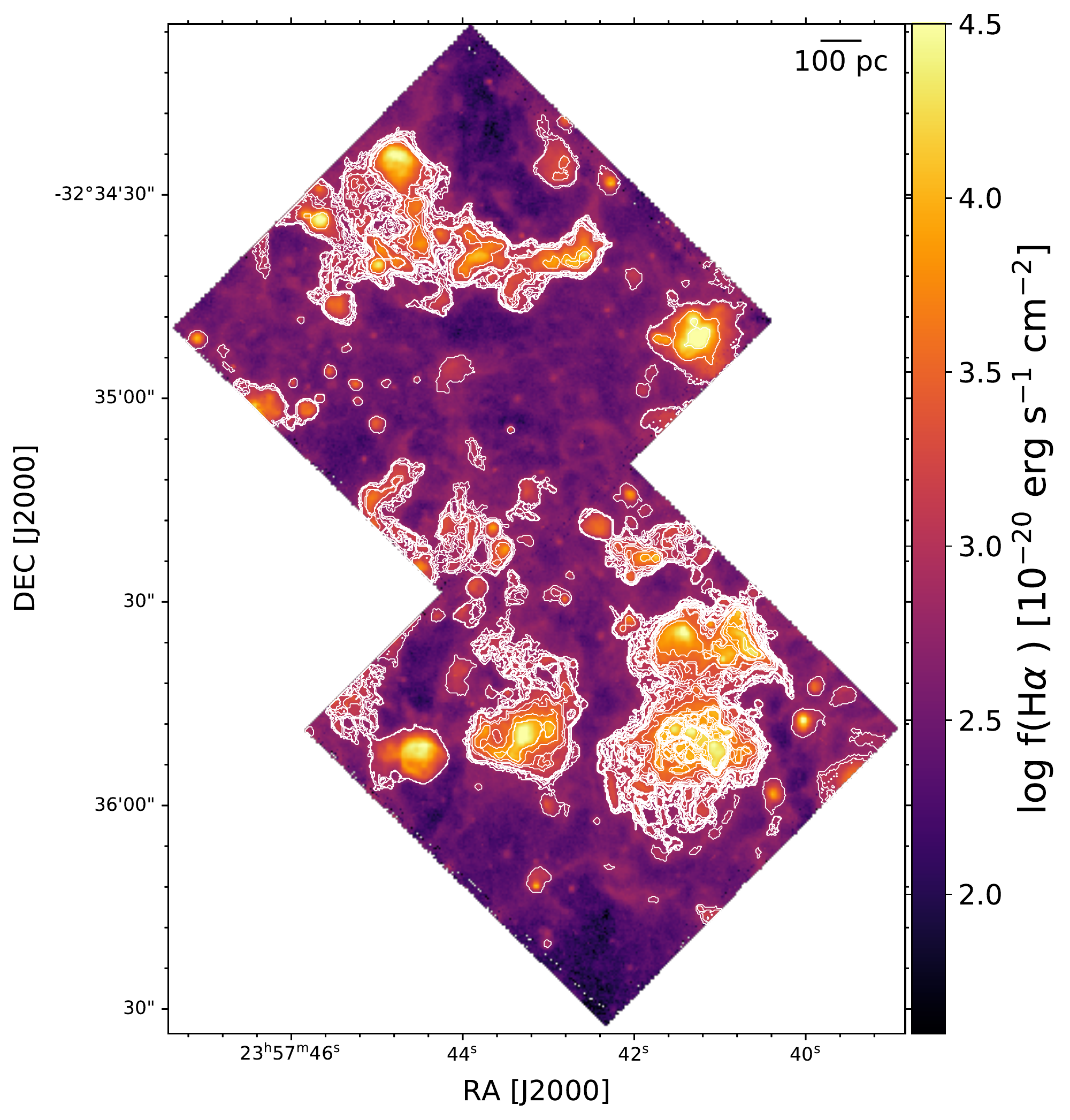}
\caption{H$\alpha$ bright regions (white contours) identified with \textsc{astrodendro} overlaid on the H$\alpha$ linemap. Single regions (leaves) are organised into a hierarchical tree consisting of various branches.}
\label{fig:dendrogram_halpha}
\end{figure}

\begin{figure*}
%\center
\includegraphics[width=5cm]{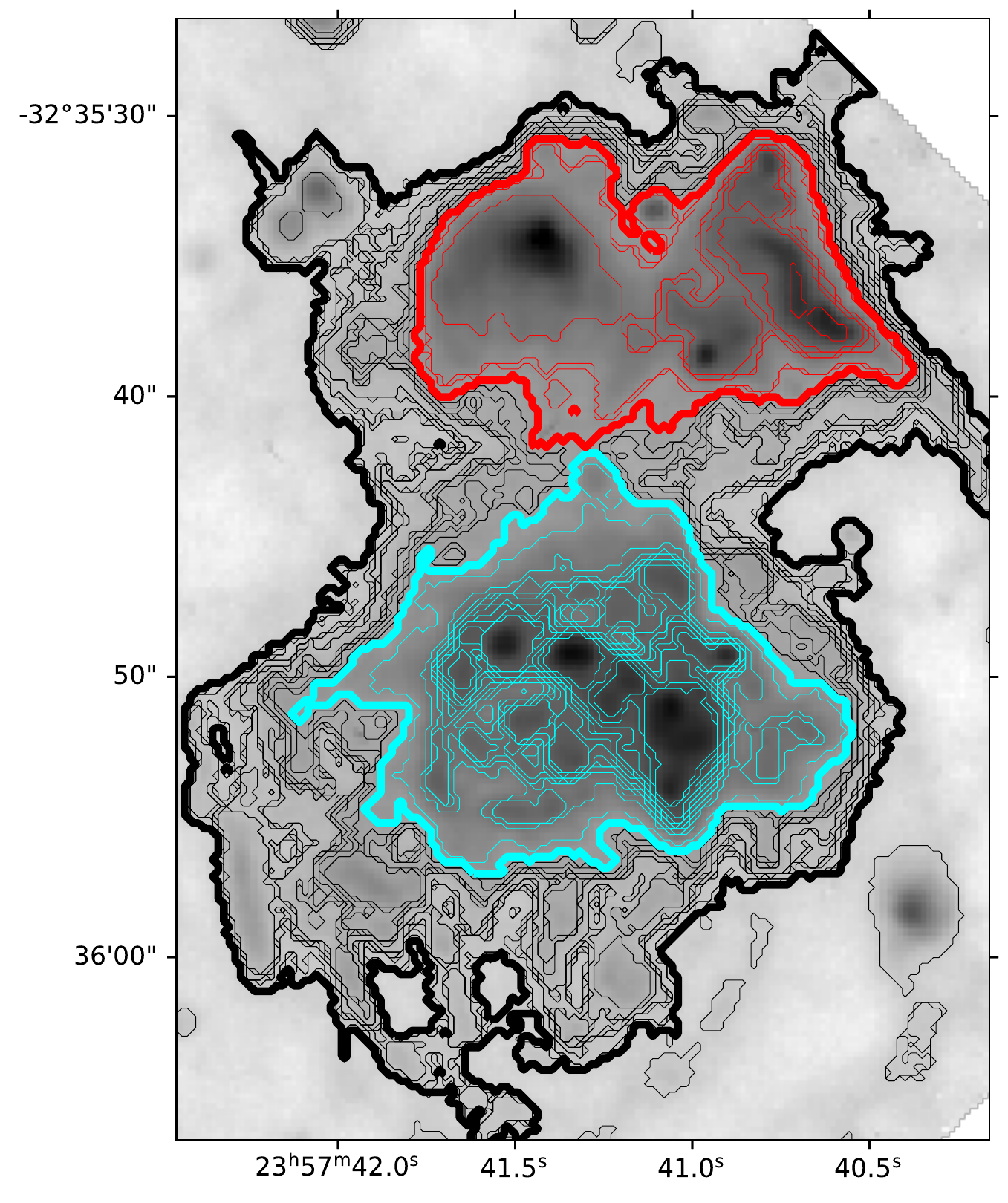}
\includegraphics[width=10cm]{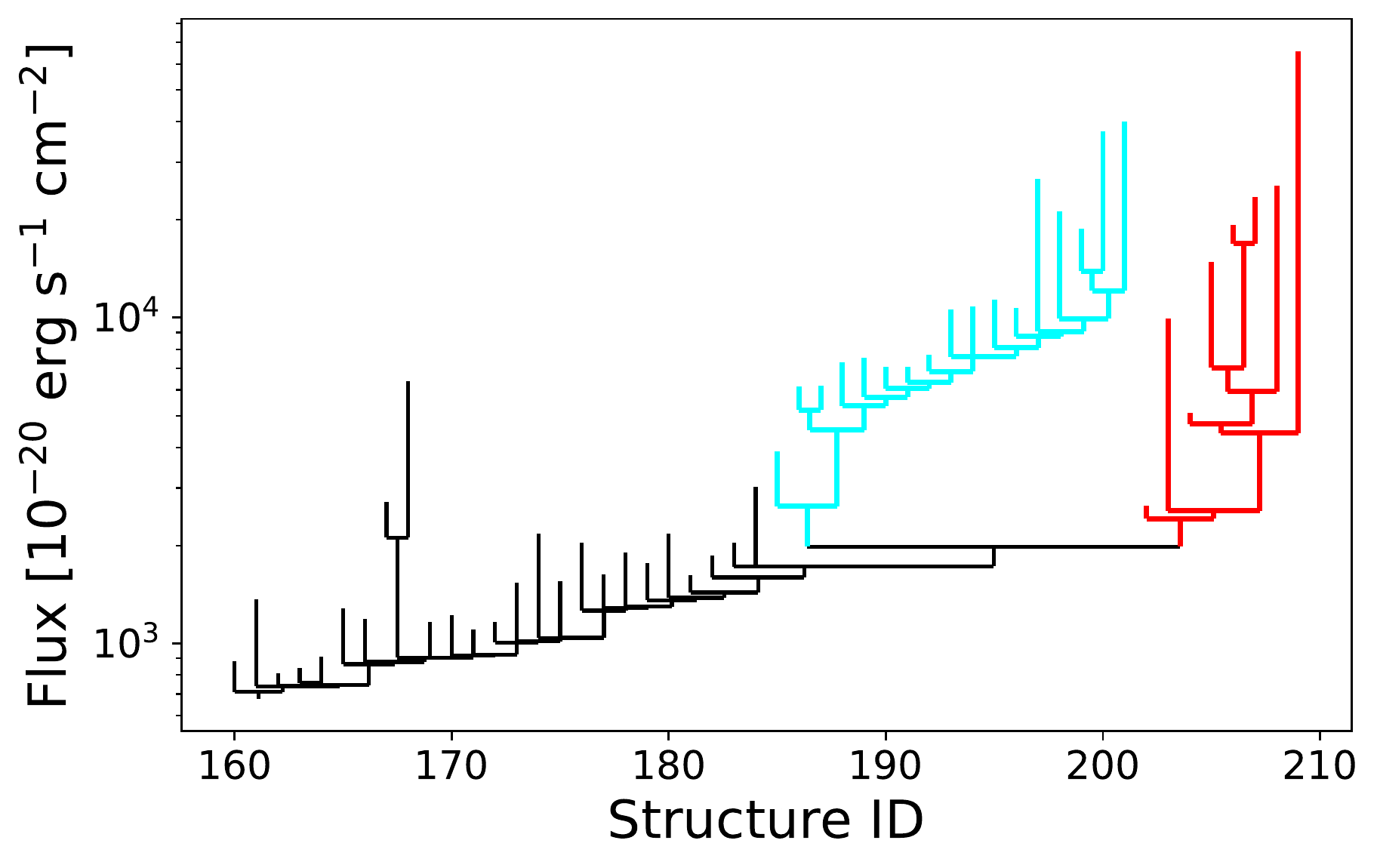}
\caption{Example of a tree structure constructed with \textsc{astrodendro} on an HII region complex. The left panel shows the structure and the right panel the corresponding tree. Highlighted in cyan and red are the two main distinct branches (`HII regions') and their respective sub-structures.}
\label{fig:tree_diagram}
\end{figure*}

%%%
In the literature a variety of approaches are used to separate HII regions from DIG, relying mainly on the properties of the H$\alpha$ emission. Widely used are \textsc{hiiphot}, a method proposed by~\citet{thilker00}, based on the gradient of the H$\alpha$ surface brightness, and the selection criterion proposed by~\citet{wang98}, which defines the boundary between DIG and HII regions according to the ratio between the local and the mean (at half-light ratio) H$\alpha$ luminosity.
We adopt a slightly different approach, and base our selection on the H$\alpha$ surface brightness and on characteristic emission line ratios, similarly to what presented by~\citet{kreckel16}. We start by selecting H$\alpha$ bright regions using the Python package \textsc{astrodendro}\footnote{\url{https://dendrograms.readthedocs.io}}. This tool allows to organise data into a hierarchical tree structure. We input the H$\alpha$ map obtained by integrating the gas cube in the rest frame wavelength range 6559 - 6568~\AA{}. To determine the depth of the tree, we assume a Gaussian shape for the diffuse radiation and estimate its median $m$ and standard deviation $\sigma$ directly on the map via sigma clipping. 
We compute the dendrogram down to $m + 10 \sigma$ (corresponding to a flux of $\SI{6.7e-18}{\erg \per \second \per \centi \metre \squared}$ spaxel$^{-1}$) and require a leaf to comprise a minimum of 15 pixels, just above the resolution limit. Various limits have been tested for the depth of the dendrogram, in the range $(m + 5 \sigma, m + 20 \sigma)$; the optimal value has been determined by visual inspection. Our cutoff value corresponds to an emission measure\footnote{The emission measure (EM) is defined as EM 
$:= \int_0^L n_e^2 dl$, where $n_e$ is the electron density and $L$ is the total path length of ionised gas. For a temperature $T = 10^4$ K, the relation between EM and H$\alpha$ surface brightness is given by: EM [pc cm$^{-6}$] = \num{4.908e17} $I_{H\alpha}$ \si{\erg \per \second \per \centi \metre \squared} arsec$^{-2}$ \citep{zurita00}.} (EM) of 82 pc cm$^{-6}$. For reference, \citet{ferguson96} used a very similar limiting value of 80 pc cm$^{-6}$ to separate DIG from HII regions in the study of NGC 7793 and NGC 247, whereas~\citet{hoopes96} used a fainter DIG limit of 50 pc cm$^{-6}$ in the sculptor galaxies NGC 55, NGC 253, and NGC 300.
This selection method results in a sample of bright regions (leaves) organised into structures (branches), as shown in Fig.~\ref{fig:dendrogram_halpha}. In Fig.~\ref{fig:tree_diagram} we show an example region (left panel) and the corresponding dendrogram (right panel). The dendrogram consists of a main branch (thick black boundary) and several sub-branches and leaves (thin black lines). The main branch splits in two main sub-branches, highlighted in red and cyan. In the following, we refer to the main branches as `HII region complexes', to their upper distinct sub-branches as 'HII regions' and to leaves as `peaks' inside the HII regions.

\begin{figure*}
\center
\includegraphics[width=18cm]{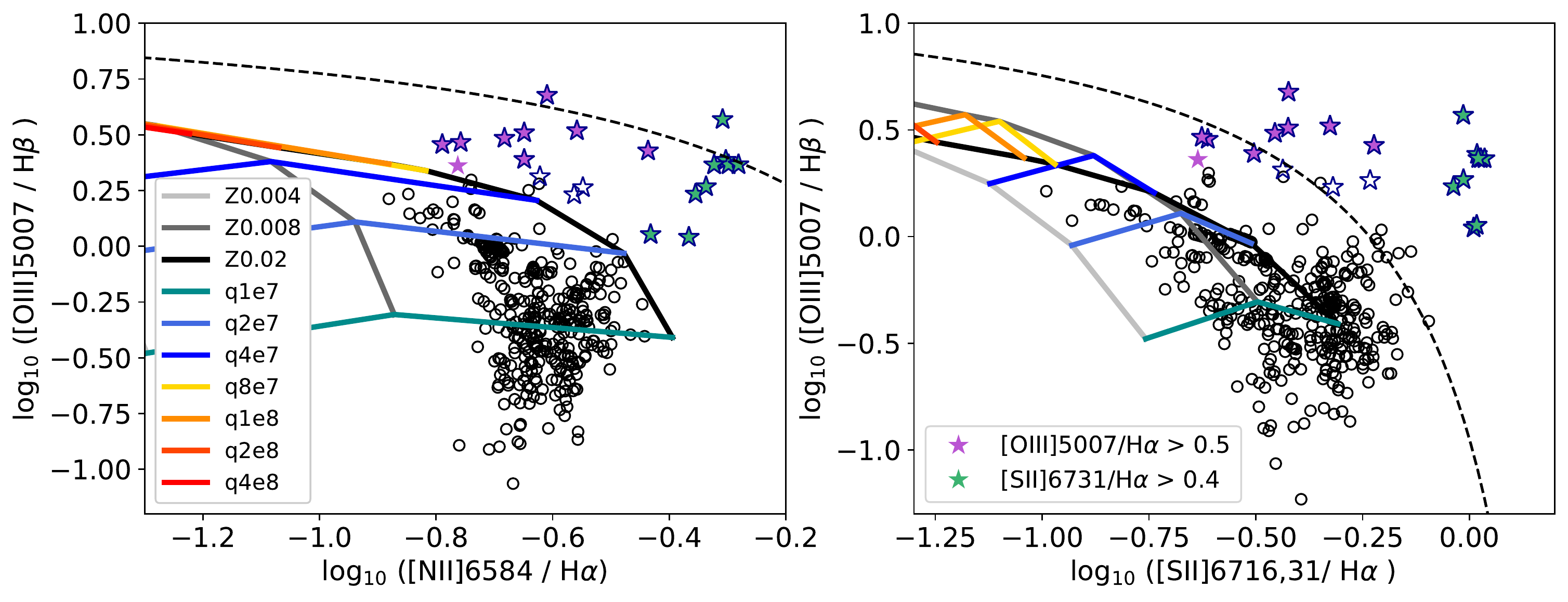}
\caption{Location of the H$\alpha$ bright regions illustrated in Fig~\ref{fig:dendrogram_halpha} on BPT diagnostic diagrams. The coloured grid indicates the models for low metallicity SF galaxies from \citet{levesque10}  (0 Myr instantaneous burst with $n_e = \SI{100}{\centi \metre}^{-3}$); the black dashed line shows for reference the extreme starburst line from \citet{kewley01}. Circles and stars correspond to regions in Fig.~\ref{fig:dendrogram_halpha}; blue stars indicate regions located > 0.1 dex from the Levesque model grid in both diagrams; filled stars indicate regions with a high ratio of [OIII]5007/H$\alpha$ (orchid) or [SII]6731/H$\alpha$ (green).}
\label{fig:bpt_hiiregions}
\end{figure*}

Selecting structures based on their brightess, however, does not provide information about the ionisation state of the gas.
In order to exclude potential contaminants, we therefore inspect the location of dendrogram structures in the `BPT' line ratio diagrams [OIII]5007/H$\beta$ vs [NII]6548/H$\alpha$ and [OIII]5007/H$\beta$ vs [SII]/H$\alpha$ \citep{baldwin81}. This type of diagram was initially devised as a way to separate star forming galaxies from objects primarily excited by other mechanisms (e.g. AGNs), but is also useful to investigate the ionisation state of resolved ISM gas.
It is sensitive to parameters such as the hardness of the radiation field, the electron density and the metallicity of the gas~\citep[see e.g.][]{kewley06}. 
The resulting diagrams are shown in Fig.~\ref{fig:bpt_hiiregions}. Each circle represents detected leaves and branches, whose flux is the total flux of the enclosed spaxels obtained from Gaussian fitting to the line. We would like to point out that also all sub-structures of the tree and not only main branches are included. We compare the position of the regions with models of low metallicity star forming galaxies from \cite{levesque10}. We adopt the instantaneous burst model grids with an electron density $n_e = \SI{100}{\centi \metre}^{-3}$, in the typical range for HII regions~\citep[see e.g.][]{osterbrock}. The models cover the metallicity range $Z = 0.001 - 0.04$ and ionisation parameter range q = \SI{1e7} - \SI{4e8}{\centi \metre \per \second}. We observe that the full range of models, with ages ranging from 0 to 5 Myr, overlaps at least in part with the data, suggesting that an age range is present in the powering sources of the HII regions. We consider as upper limit the 0 Myr grid shown in Fig.~\ref{fig:bpt_hiiregions}, which also follows best the distribution of the observed points. We flag structures lying > 0.1 dex above the grid in both BPT diagrams (marked as blue stars in Fig.~\ref{fig:bpt_hiiregions}); these regions are analysed in Sect.~\ref{section:outliers}. For comparison, the black dashed line indicates the `extreme starburst line' from \citet{kewley01}. We observe that our data are better represented by slightly sub-solar metallicity models, and by a ionisation parameter q < \SI{4e7}{\centi \metre \per \second}.

\par
\begin{figure}
\center
\includegraphics[width=9cm]{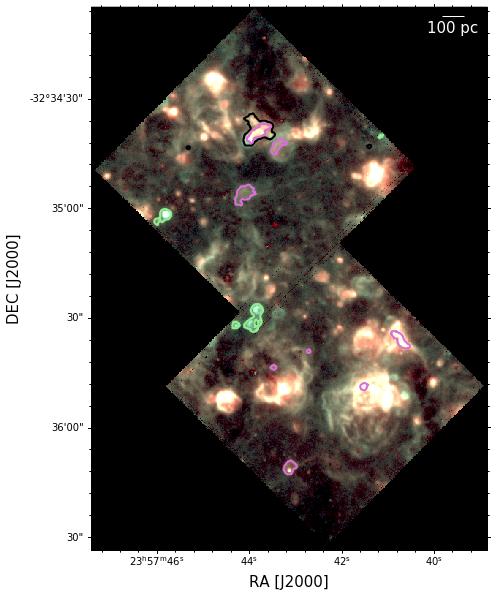}
\caption{Spatial location of the H$\alpha$ bright regions that show inconsistency with models of star forming galaxies (in black; star symbols in Fig.~\ref{fig:bpt_hiiregions}) or having a ratio [OIII]5007/H$\alpha$ > 0.5 or [SII]6731/H$\alpha$ > 0.4 (orchid and green respectively; filled stars in Fig.~\ref{fig:bpt_hiiregions}) overlaid on an RGB image of the gas emission (R: H$\alpha$, G: [SII]6716,6731, B: [OIII]4959,5007).}
\label{fig:rgb_outliers}
\end{figure}

\begin{figure*}
%\center
\includegraphics[width=9cm]{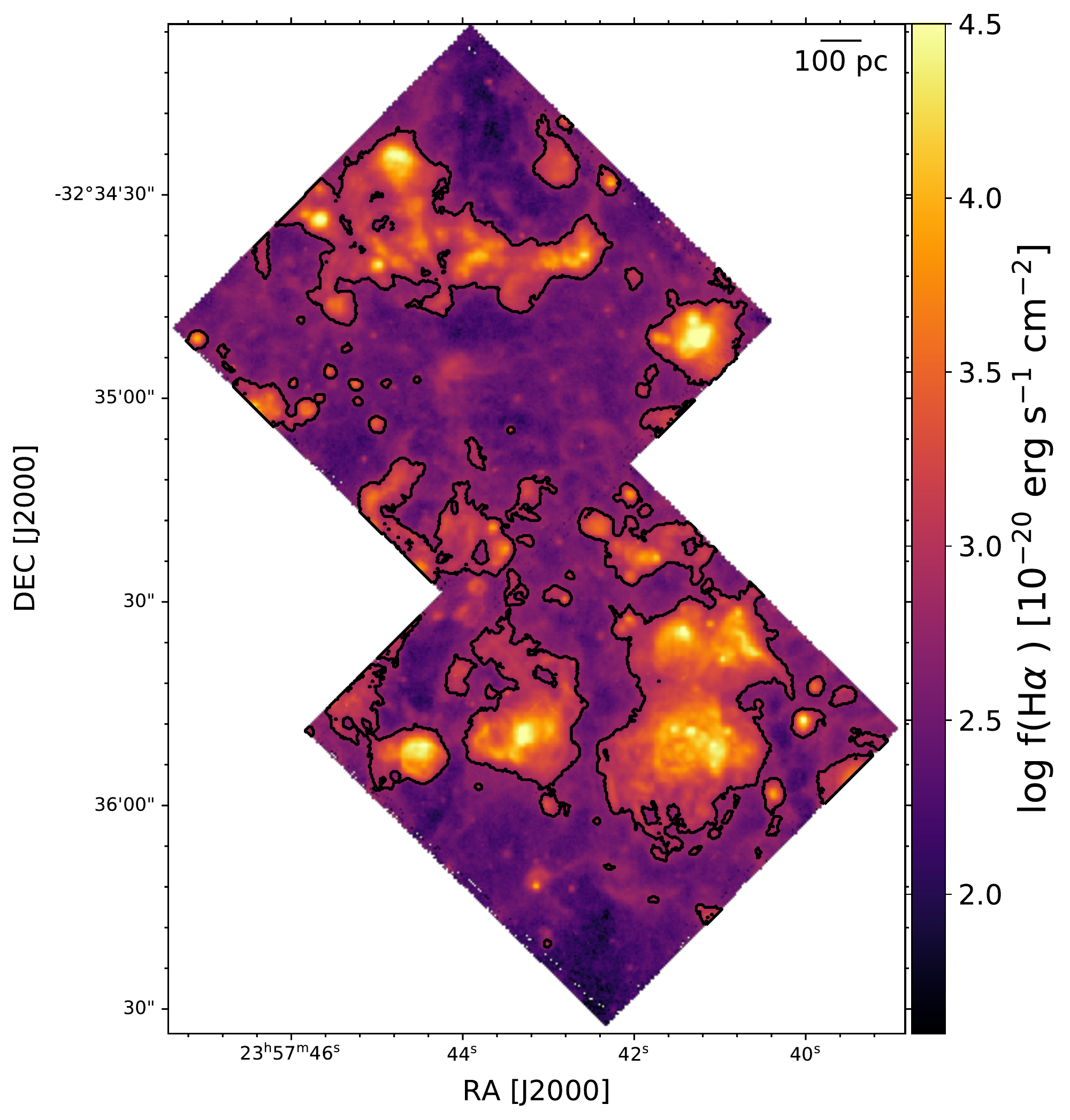}
\includegraphics[width=9cm]{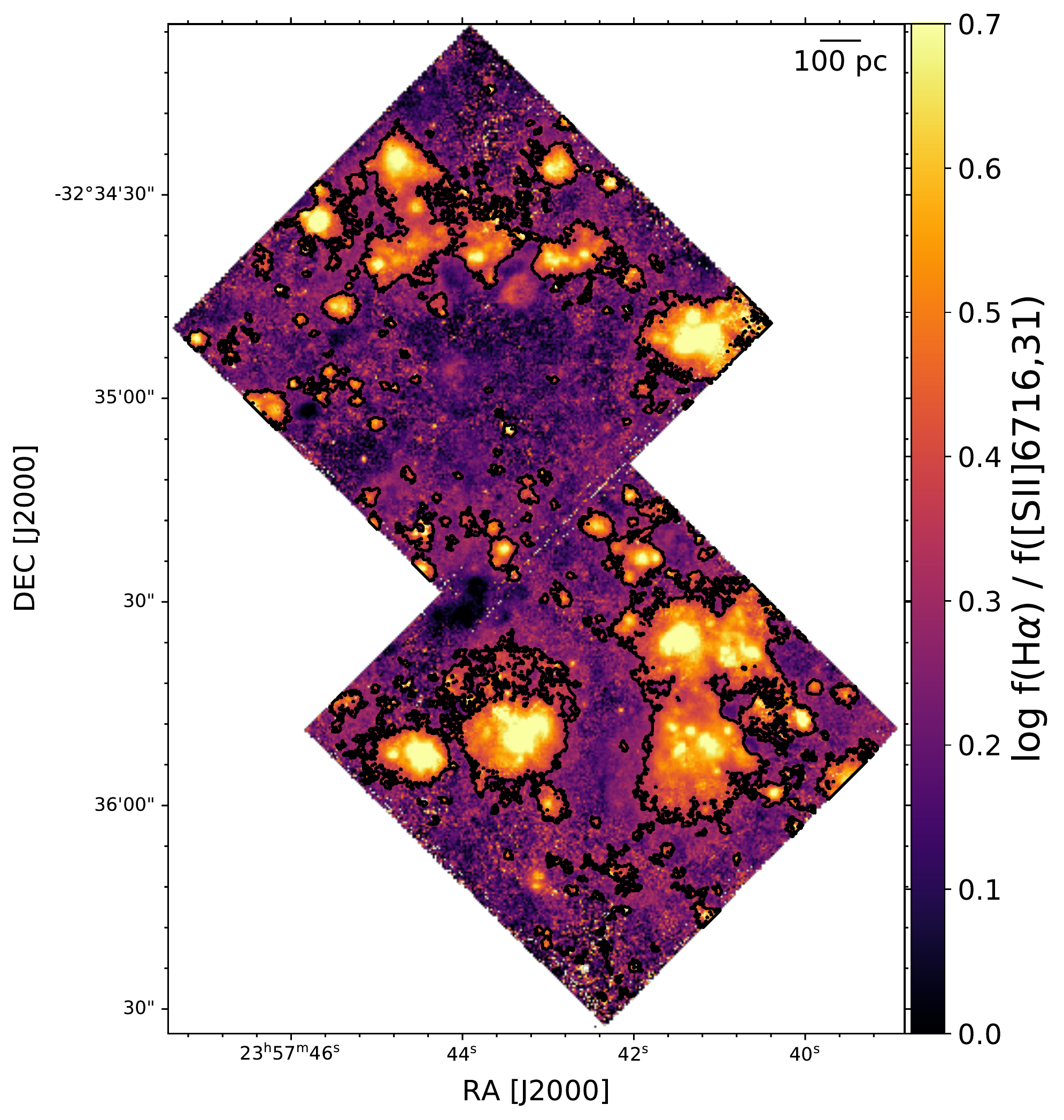}
\includegraphics[width=9cm]{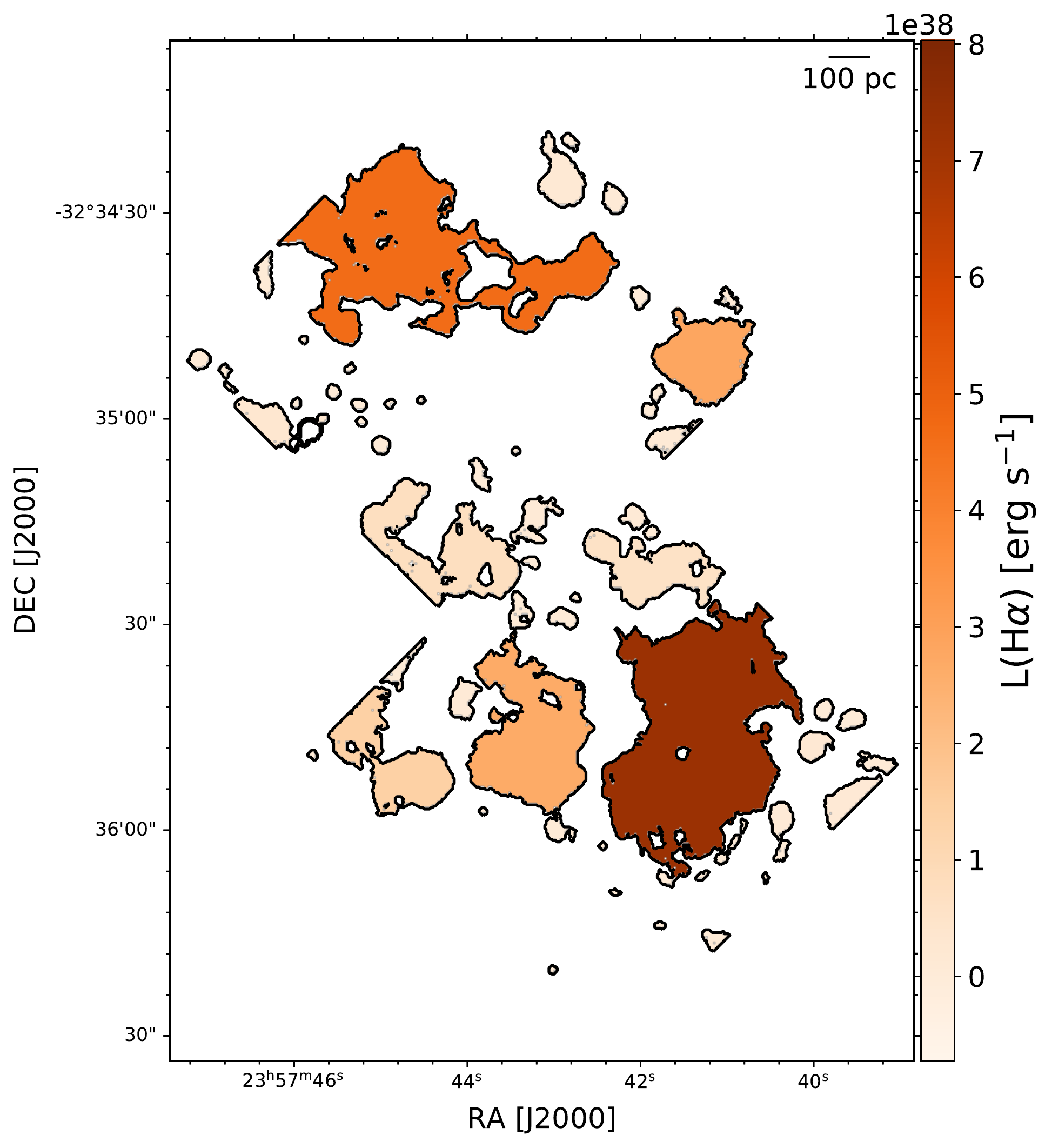}
\includegraphics[width=9cm]{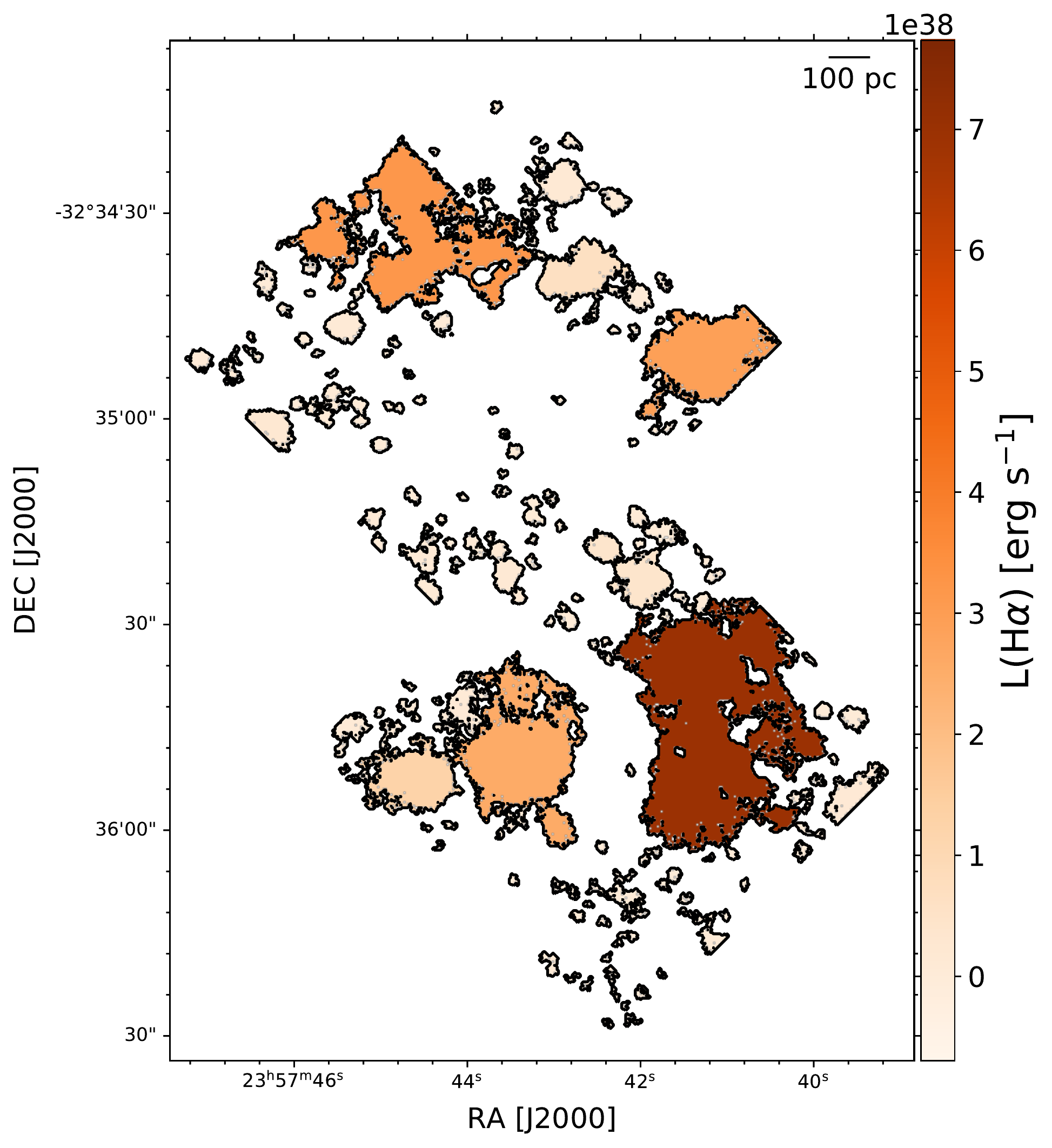}
\caption{HII regions samples constructed on the H$\alpha$ map (left) and H$\alpha$/[SII] map (right). Top: outermost regions contours (black) overlaid on the linemap; bottom: total luminosity of the regions. One can notice that the H$\alpha$/[SII]-selection results in a more conservative sample.}
\label{fig:hii_regions}
\end{figure*}

 We further inspect two additional line ratios: [OIII]5007/H$\alpha$, highlighting high excitation objects, such as planetary nebulae (PNe), and [SII]6731/H$\alpha$, enhanced in supernova remnants (SNR). From a sample of PNe in local galaxies, \citet{ciardullo02} inferred a typical ratio [OIII]5007/H$\alpha$ > 2, whereas
to separate shock-heated SNR nebulae from photoionised gas, the canonical value in the Milky Way and in local group galaxies is [SII]6731/H$\alpha$ > 0.4~\citep[see e.g.][]{blair97}. Since our target has a lower metallicity, both ratios are suppressed, as we can observe when comparing the \cite{levesque10} model grid with the \citet{kewley01} extreme starburst line in Fig~\ref{fig:bpt_hiiregions}. We therefore inspect all regions featuring a ratio in [OIII]/H$\alpha$ (PN candidates) and [SII]/H$\alpha$ (SNR candidates) that is above the 95 percentile of the line ratio observed over the whole FoV. Regions flagged with this method are shown as filled coloured stars in Fig.~\ref{fig:bpt_hiiregions}. We observe that almost all of these structures also stand out in the BPT analysis. The spatial location of all the flagged regions is shown in Fig.~\ref{fig:rgb_outliers}. We investigate their nature further in Sect.~\ref{section:outliers}. We note that one of the regions will result to be associated with a WR star and is kept in the HII regions sample. The resulting sample of HII region is shown in Fig. \ref{fig:hii_regions} (upper left panel). We also show the total luminosity of the HII-complexes (lower left panel), to offer a comparison with similar studies at larger distances~\citep[e.g.][]{weilbacher18}.

\subsection{Selection of PNe, SNRs and WR star candidates}
\label{section:outliers}
In Sect. \ref{section:hiiregions_selection} we have inspected the ionisation state of H$\alpha$ bright regions and flagged some showing peculiar line ratios. Here we further analyse their nature and build a sample of candidate PNe, SNRs, and WRs.

\par
\begin{figure}
\center
\includegraphics[width=9cm]{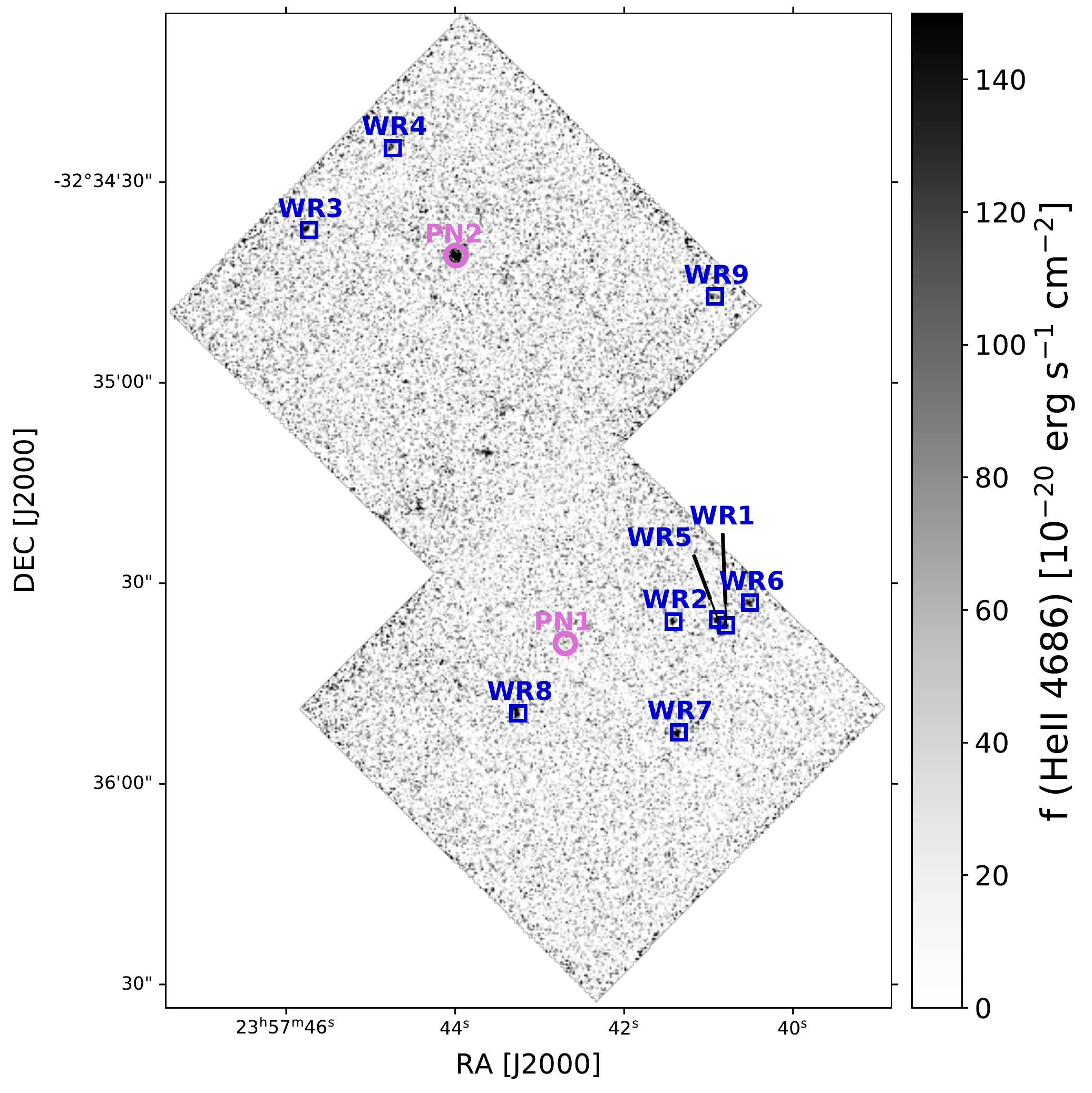}
\caption{HeII 4686 linemap obtained by integrating the full cube (gas and stars) in the rest frame wavelength range 4682 - 4691~\AA{} and manually subtracting the continuum. The location of identified WR stars is marked in blue; the one of candidate PNe in orchid.}
\label{fig:he2_linemap}
\end{figure}

\begin{figure}
\center
\includegraphics[width=9cm]{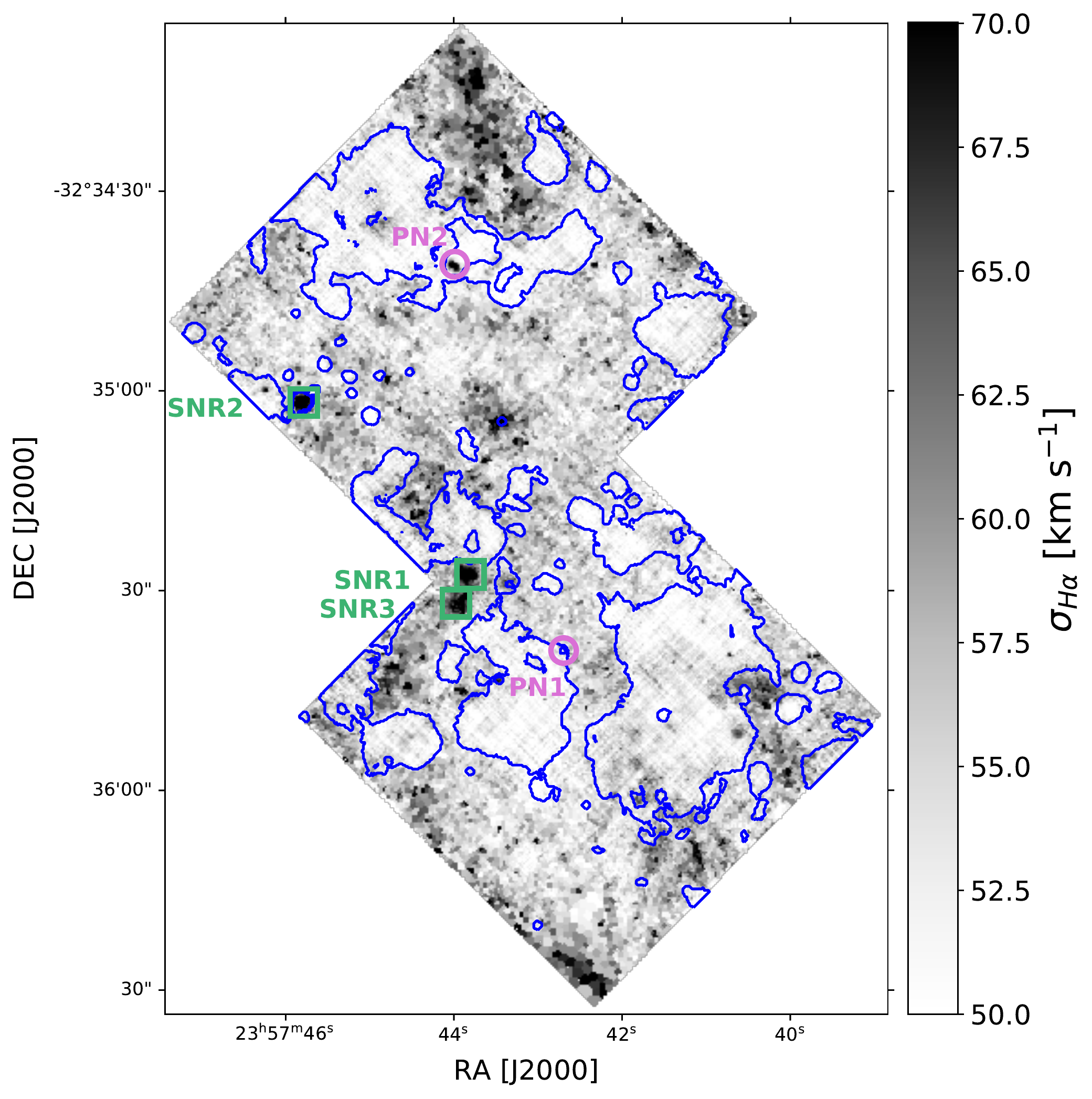}
\caption{Velocity dispersion of the H$\alpha$ line (same as Fig.~\ref{fig:halpha_kinematics}). Overlaid are the HII regions contours (blue; sample selected on the H$\alpha$ map) and the location of candidate PNe (orchid) and SNR (green).}
\label{fig:sigma_halpha_dig}
\end{figure}

 We have identified regions of high excitation based on their [OIII]5007/H$\alpha$ ratio. We inspect whether the latter also exhibit narrow HeII 4686 line emission, indicating a very strongly ionised medium, in which case we include them in a sample of candidates PNe. We identify two candidates, whose location is shown overlaid on the HeII linemap in Fig.~\ref{fig:he2_linemap}. One of the two exhibits a very strong HeII emission as well as a high H$\alpha$ velocity dispersion $\sigma \sim 85$ km/h (see Fig.~\ref{fig:sigma_halpha_dig}). We would like to point out that since PNe are not resolved at our distance, we do not necessarily expect them to stand out in a velocity dispersion map. We extract spectra of the candidate using a circular aperture of 0.4\arcsec, corresponding to the resolution limit (see Fig.~\ref{fig:pn_spectra}): on the extracted spectra we measure an apparent magnitude $m_{[OIII]5007} = 25.70$ and 23.92 respectively for candidate PN1 and PN2 in Table~\ref{table:candidates}. Our result does not change if we extract the candidates with \textsc{astrodendro} directly on the [OIII]4959,5007 linemap and with a minimum leaf size corresponding to the resolution limit.

\par We then inspect regions with a high [SII]6731/H$\alpha$ ratio to search for SNR candidates. Three of these also show a high H$\alpha$ velocity dispersion of $\sim 75-100$ km/s (Fig.~\ref{fig:sigma_halpha_dig}); we confirm these as candidate SNR. The position of the candidates is shown in Fig.~\ref{fig:sigma_halpha_dig}: two of these were previously catalogued by \citet{blair97} \citep[with position revised in][]{pannuti01}, using H$\alpha$ and [S II] narrowband filters and the requirement H$\alpha$/[SII] > 0.4.
\citet{pannuti01} also lists another optical candidate in our FoV. Although we do observe a higher line ratio [SII]6731/H$\alpha$ $\sim$ 0.24 and an enhanced gas velocity dispersion at the corresponding location, this region does not make our cut. Also in this case, results do not change if we extract the candidates with \textsc{astrodendro} directly on the [SII]6731 linemap.

\par On the HeII 4686 linemap we also extract WR star candidates. We run \textsc{astrodendro} with a minimum pixel size of 10 pixels,
corresponding to the resolution limit, and down to 40$\sigma$ above the map background (corresponding to a flux of $\sim \SI{1.0e-18}{\erg \per \second \per \centi \metre \squared}$) to extract only the brightest regions. We then confirm visually the presence of the characteristic HeII 4686 bump (see spectra in Fig.~\ref{fig:wr_spectra}). This line profile is an indication of stellar winds and clearly distinguishes WR stars from the narrow HeII 4686 emission of PNe. We note that another characteristic feature of WR stars, the CIV 5808 bump, falls inside the AO laser gap for our target.
We identify nine WR stars (see Fig.~\ref{fig:he2_linemap}), of which seven were never catalogued before. \cite{bibby10} previously identified three candidate WR in the MUSE FoV; we confirm two of these (see last two columns of Table~\ref{table:candidates}). We do not observe any HeII emission at the location of the third candidate \citep[ID 20 in][]{bibby10}.
\citet{wofford20} select candidate luminous blue variable (LBV) stars combining the LEGUS-HST data with the MUSE dataset presented in this work and independently identify four of the candidates WRs presented here.
We see that one of the H$\alpha$ bright regions flagged in Sect.~\ref{section:hiiregions_selection} due to its high [OIII]5007/H$\alpha$ ratio is associated with a WR star.

\par WR, SNR and PNe candidates are listed in Table~\ref{table:candidates}. Spectra of all the candidates can be found in the Appendix (Fig.~\ref{fig:pn_spectra}, \ref{fig:snr_spectra} and \ref{fig:wr_spectra}). We do not identify the origin of the remaining regions highlighted in Fig.~\ref{fig:rgb_outliers}, and we label them as `regions ionised by other mechanisms' in the rest of our analysis, as they exhibit features typical of shocked regions.

\begin{table}
\caption{SNRs, PNe and WRs identified in the MUSE FoV and reference to previous detections. Spectra of all candidates can be found in the Appendix.}
\begin{tabular}{ccccc}
\hline \hline
Obj. ID & Ra (J2000) & Dec (J2000) & Ref. & Ref. ID \\ \hline
SNR1 & 23:57:43.82 & -32:35:27.8 & 1 & N7793-S6 \\
SNR2 & 23:57:45.80 & -32:35:01.9 & 1 & N7793-S10 \\
SNR3 & 23:57:44.00 & -32:35:32.1 &  & \\
PN1 & 23:57:42.71 & -32:35:39.2 & \\
PN2 & 23:57:44.01 & -32:34:41.2 & \\
WR1 & 23:57:40.81 & -32:35:36.5 &  2  & 10 (WN2-4b) \\
WR2 & 23:57:41.43 & -32:35:35.9 &  2  & 14 (WC4) \\  
WR3 & 23:57:45.74 &-32:34:37.4 & \\
WR4 & 23:57:44.75 &-32:34:25.1 & \\
WR5 & 23:57:40.90 &-32:35:35.6 & \\
WR6 & 23:57:40.53 & -32:35:33.1 & & \\
WR7 & 23:57:41.37 &-32:35:52.5 &  & \\
WR8 & 23:57:43.27 & -32:35:49.6 & \\
WR9 & 23:57:40.94 &-32:34:47.3  & \\
\hline
\end{tabular}
\tablebib{(1) \citet{pannuti01}; (2) \citet{bibby10}}
\label{table:candidates}
\end{table}

\subsection{Testing different ways to select HII regions}
\label{section:hii_selection_2}
As a way of comparison, we perform a different selection procedure to extract HII region boundaries, using the line ratio map H$\alpha$/[SII], as in~\citet{kreckel16}. Such analysis is only possible thanks to the fact that we detect both lines with a sufficiently high S/N across the whole FoV (H$\alpha$ with S/N > 5 and the combined [SII]6716,6731 lines with S/N > 3). This ratio is sensitive to the ionisation state of the gas, due to the slightly lower ionisation potential of SII (10.4 eV vs 13.6 eV for HII), and allows us to define better physically motivated HII regions boundaries. \citet{pellegrini12} adopts a similar approach, combining H$\alpha$ emission and the [SII]/[OIII] ratio, to identify HII regions in the LMC and SMC. This approach proves to be effective especially in the case of a density bounded region. We use the same H$\alpha$ map as in Sect.~\ref{section:hiiregions_selection} and create a high resolution [SII] map by integrating the gas cube in the rest frame range 6713 - 6721 and 6728 - 6735~\AA{}.
We construct a dendrogram on the H$\alpha$/[SII] linemap to a depth of 5 $\sigma$ above the diffuse background mean (corresponding to a line ratio of $2.1$) and then follow the same steps described in Sect.~\ref{section:hiiregions_selection} to obtain a similar HII regions sample. The resulting regions are shown in Fig.~\ref{fig:hii_regions} (upper right panel). We see that this approach results in a more conservative sample: the boundaries of the regions are moved to include only gas that exhibits a higher degree of ionisation. This yields, therefore, a larger value for the DIG contribution to the total H$\alpha$ luminosity within our FoV: we discuss this point in Sect.~\ref{section:dig_properties}. For the remainder of this work we use the regions selected on the H$\alpha$ map as a reference.

\section{Electron temperature and density}
\label{section:ne_Te}
\begin{figure}
\center
\includegraphics[width=7.8cm]{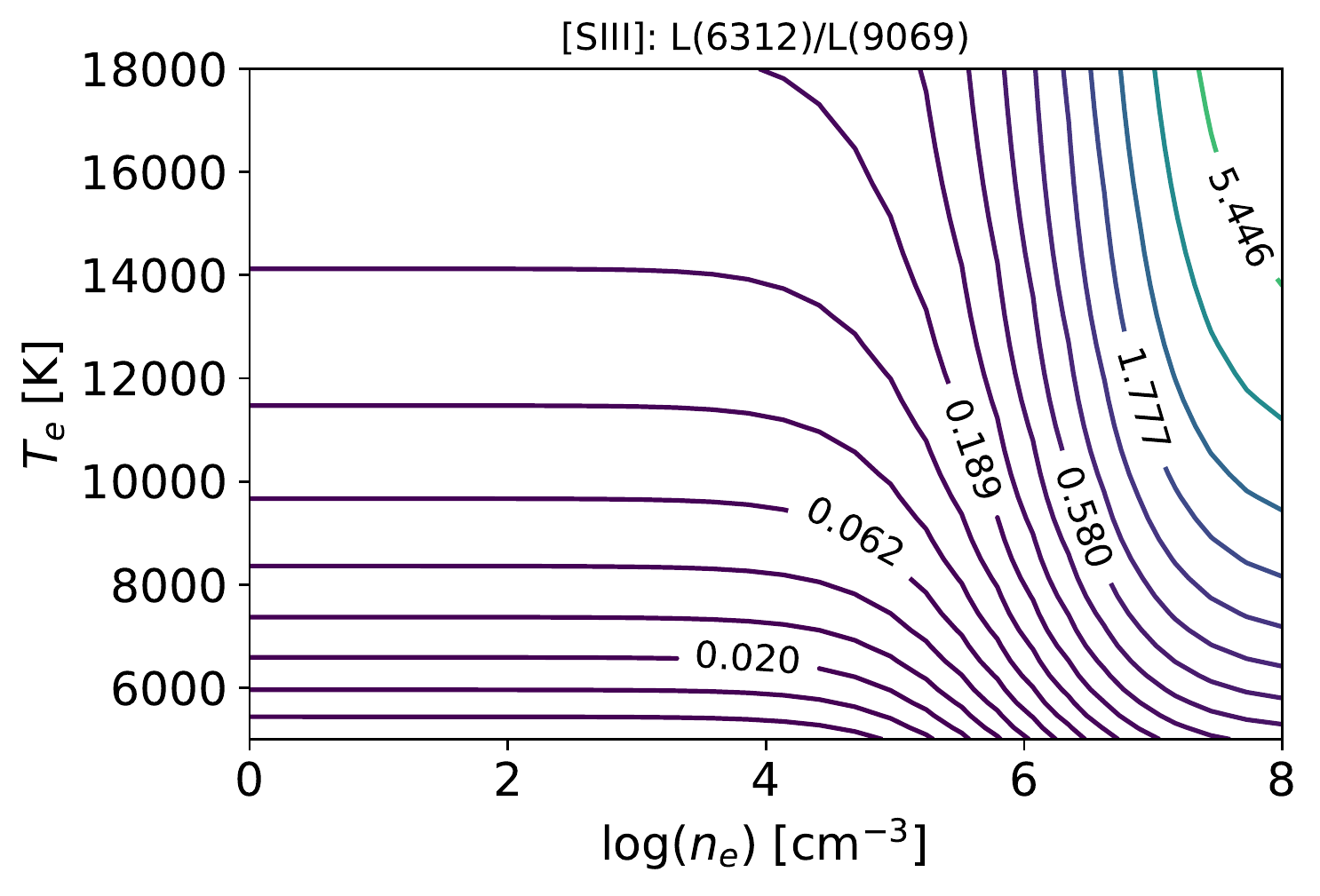}
\includegraphics[width=8.1cm]{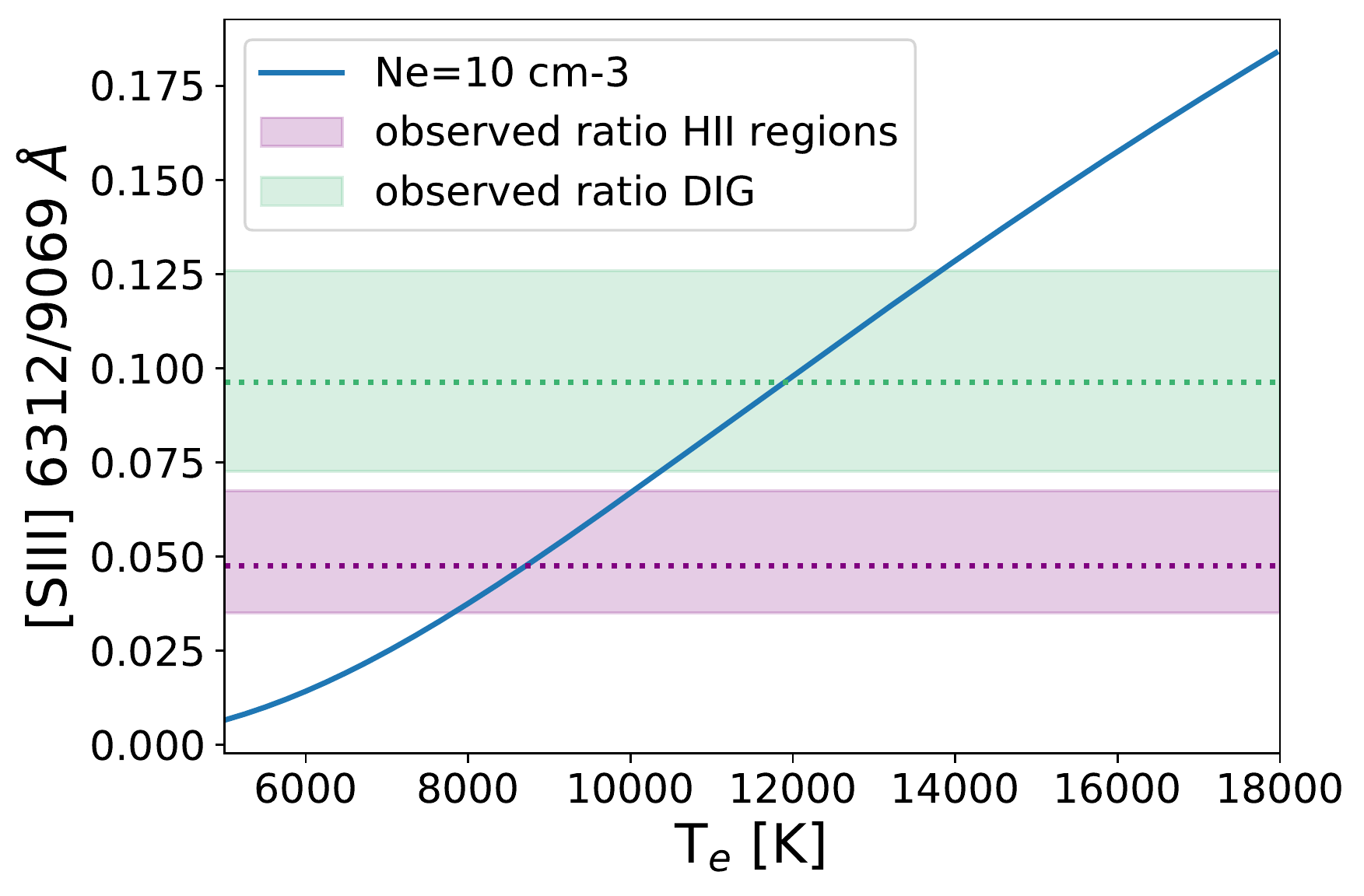}
\includegraphics[width=7.6cm]{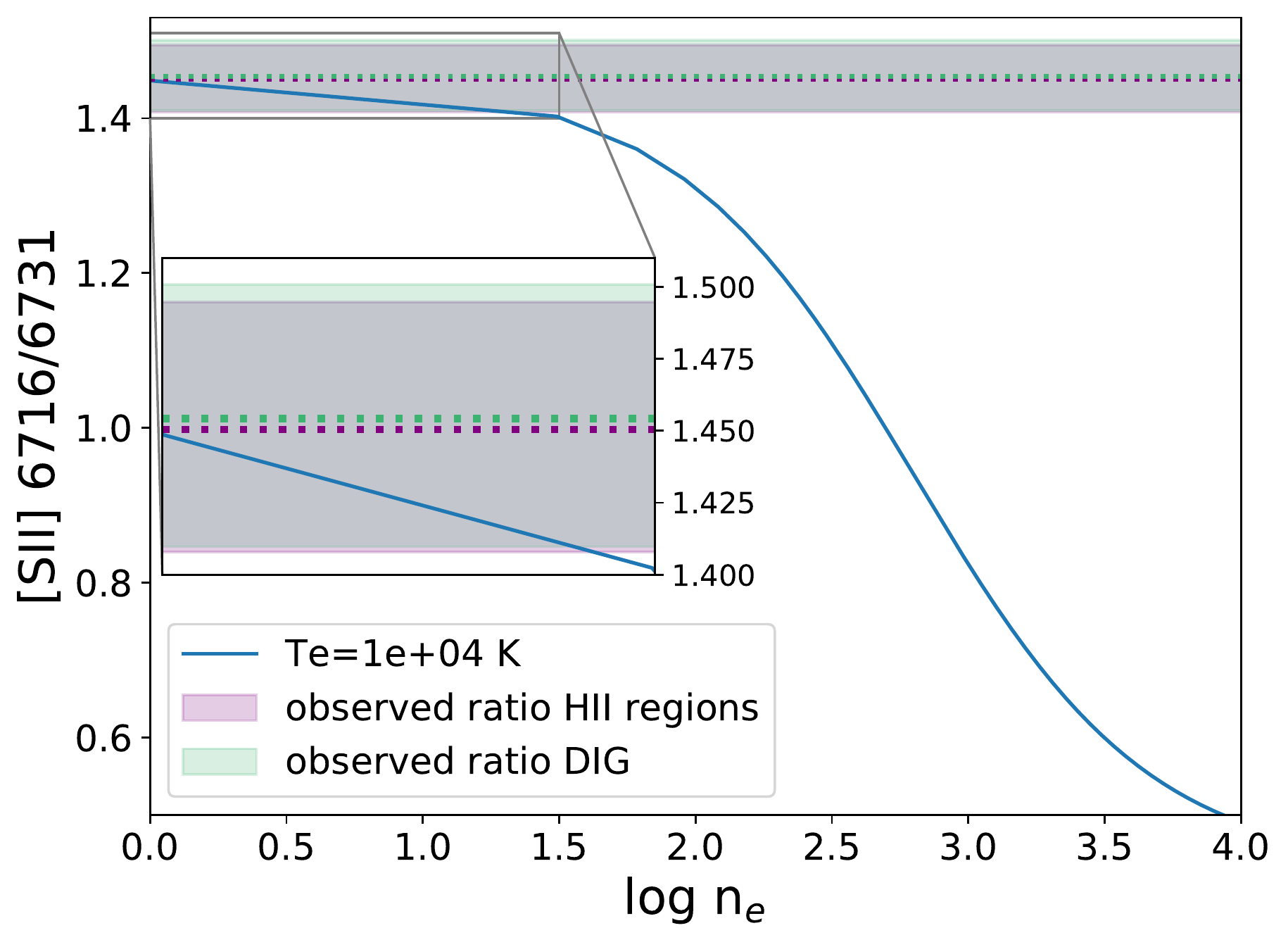}
\caption{Top: Emissivity grid for the [SIII]6312/9069 line ratio, from \textsc{pyneb}. For densities $n_e$ below the critical value $\log n_e \lessapprox 3$, the temperature is almost independent from the density. Centre and bottom: dependence of the [SIII]6312/9069 line ratio on the electron temperature and of the [SII]6716/6731 line ratio on the electron density, for a fixed value of the second quantity. Shaded areas indicate the range of observed values for the HII regions (purple) and DIG (green). The coloured area spans the first to third quartile of the distribution; the median is indicated as a dotted line. The method is unable to recover densities below 1 cm$^{-3}$ (corresponding to a line ratio [SII]6716/6731 $\lesssim$ 1.45).}
\label{fig:emisgrid_s3}
\end{figure}

\begin{figure*}
%\center
\includegraphics[width=9cm]{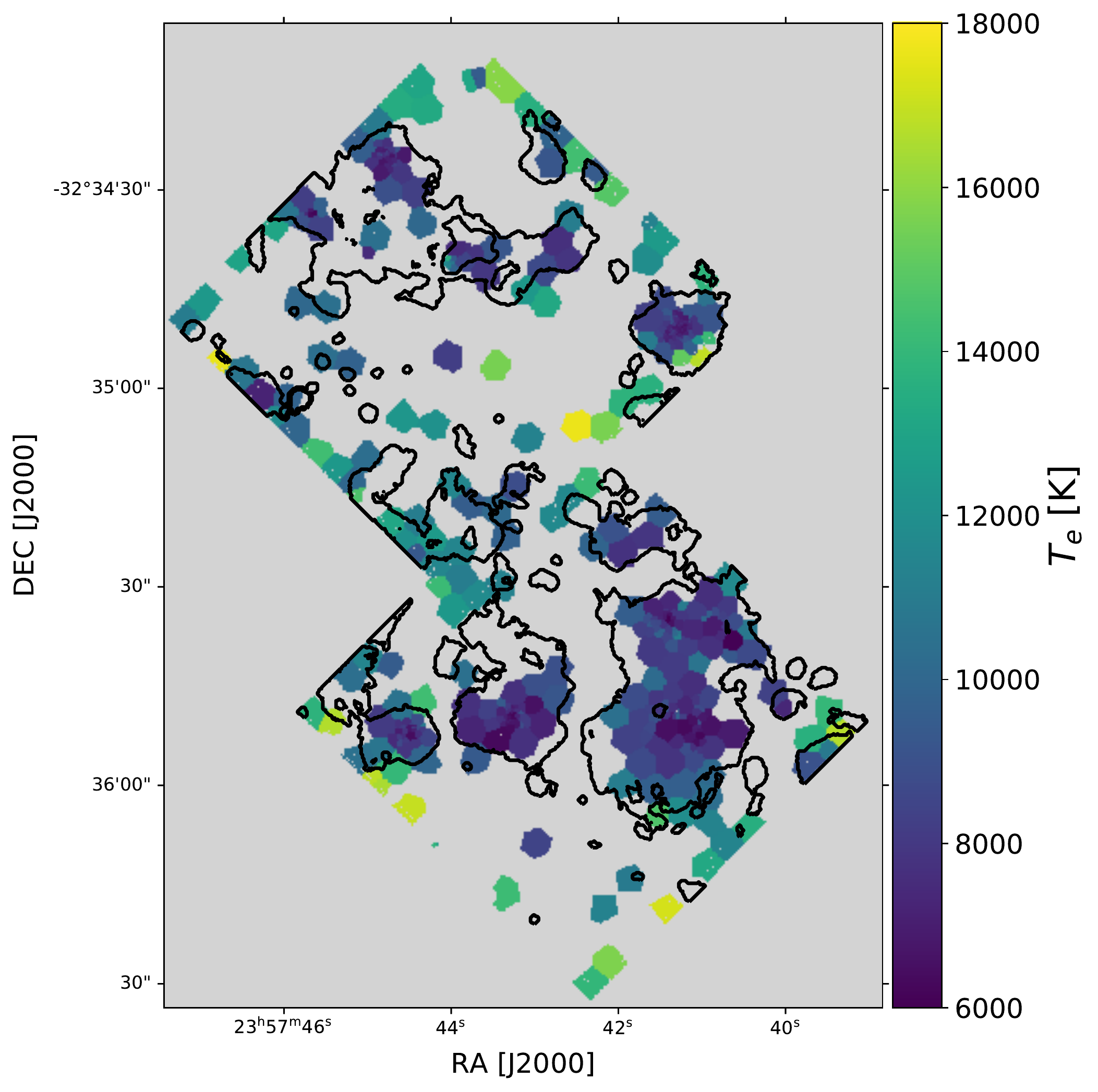}
\includegraphics[width=8.5cm]{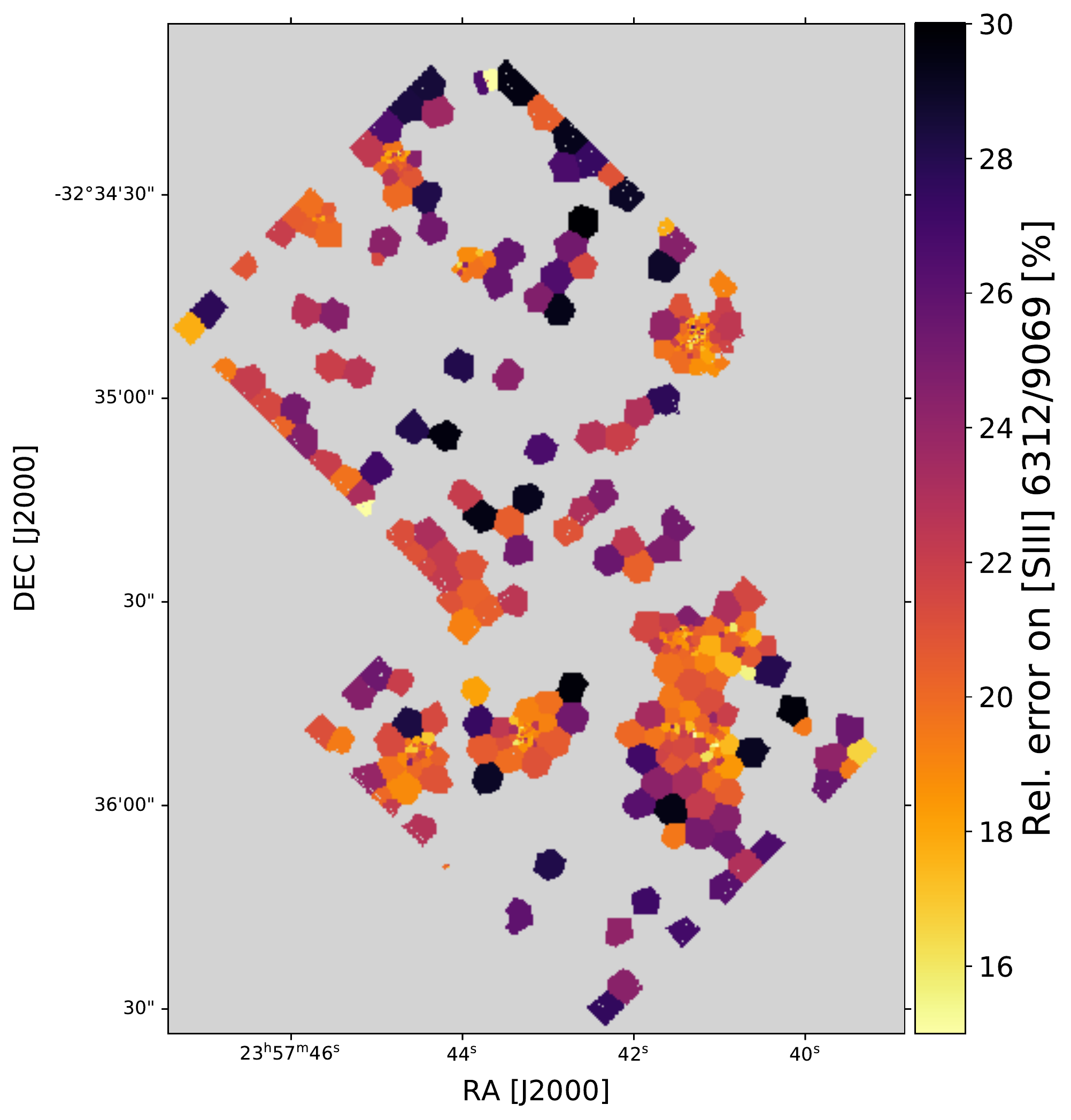}
\caption{Left: electron temperature computed with \textsc{pyneb}, for an assumed density $n_e = \SI{10}{\per \cubic \centi \metre}$. HII region contours are overlaid in black. Right: relative error on the [SIII]6312/9069 line ratio. Grey-shaded areas indicate bins where either the [SIII]6312 line was not detected, or the error on the [SIII] line ratio is > 30\%.}
\label{fig:Te}
\end{figure*}

We determine the electron temperature ($T_e$) and density ($n_e$) of the ionised gas across our FoV with \textsc{Pyneb}~\citep{Luridiana15}.
The relation between the ratio of [SIII]6312/9069 and the temperature is shown in Fig.~\ref{fig:emisgrid_s3} (first two panels); it is apparent that the values of $n_e$ relevant to our study ($n_e << \SI{1000}{\per \cubic \centi \metre}$) are in the regime where every collisional excitation if followed by spontaneous radiative decay and the temperature is independent from the density. For the temperature determination we assume a fiducial constant density value $n_e \sim \SI{10}{\per \cubic \centi \metre}$. We obtain maps of the two [SIII] lines by integrating respectively the gas cube in the rest frame wavelength range 6310 - 6317~\AA{} and a continuum subtracted cube obtained by extrapolating the best \textsc{ppxf} solution beyond the range of the fit in the wavelength range 9064 - 9077~\AA{}. We tessellate the resulting ratio map to a S/N $\sim$ 3 in [SIII]6312. We then consider only the Voronoi bins in which the relative error on the [SIII] line ratio is below 30 \% (see the right panel of Fig.~\ref{fig:Te}). The main limitation, however, lies in the detection of the weak [SIII]6312 line. Fig.~\ref{fig:Te} (left panel) shows the resulting temperature map. 
For each Voronoi bin we run \textsc{pyneb} on 1000 Montecarlo realisations of the line ratio, sampling a Gaussian distribution with width equal to the 1$\sigma$ error on the ratio. We estimate an uncertainty on the temperature in each Voronoi bin adopting as upper and lower limit the 14th and 86th percentile of the resulting temperature distribution.

\par
\begin{figure}
\center
\includegraphics[width=9cm]{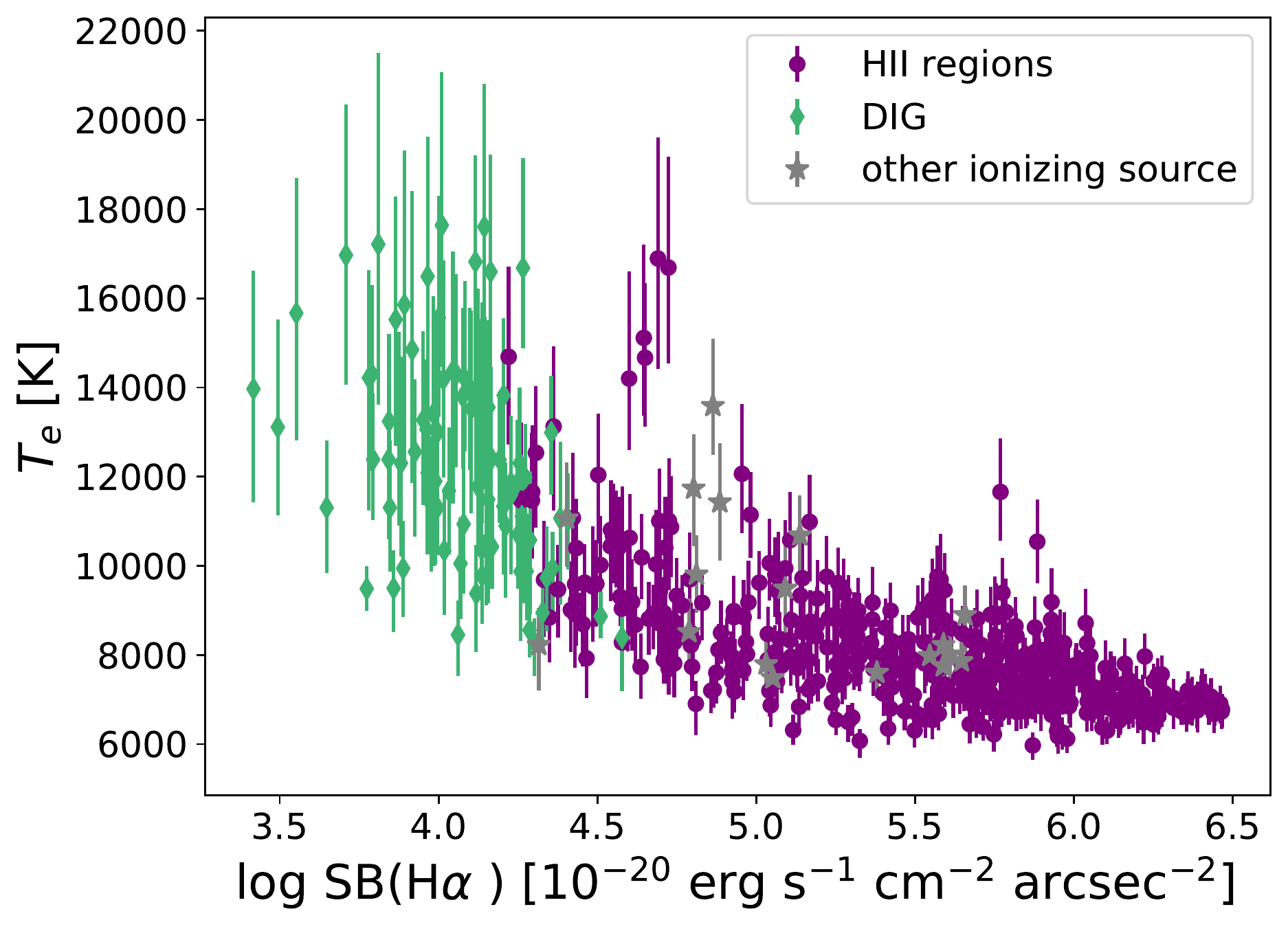}
\caption{Electron temperature as function of H$\alpha$ surface brightness for each Voronoi bin in Fig.~\ref{fig:Te}. Points are labelled according to whether the corresponding Voronoi bin is part of an HII region (purple circles), the DIG (green diamonds) or of a region ionised by another source (grey stars). Bins with a relative error on the [SIII] line ratio > 30\% have been excluded, as wells as bins with $T_e > 18000$ (spurious detections on the edge of the FoV).}
\label{fig:Te_vs_halpha}
\end{figure}

 We compare the temperature map to the location of the HII regions and recover a temperature $T_{e} = 8714_{-903}^{+1319}$ K in HII regions
and
$T_{e} = 12087_{-1718}^{+1880}$ K in the DIG,
where we have indicated the median and the first and third quartile of the distribution. Hereby we have excluded bins with unrealistically high temperatures $T_e > 18000$ K, all caused by spurious detections on the edge of the field. This is in line with observations of the Milky Way, where several studies have shown that the temperature of the diffuse gas is $\sim$ 2000 K higher than in HII regions~\citep[see e.g.][]{haffner09,madsen06}. In general, we observe a larger spread towards high temperatures in lower surface brightness regions. This seems to indicate a trend of increasing temperature with decreasing surface brightness, as shown in Fig.~\ref{fig:Te_vs_halpha}. However, since we are mainly limited by the ability to detect the weaker [SIII]6312 line, we cannot rule out the existence of a population of lower H$\alpha$ surface brightness regions at low $T_e$.

\par
\begin{figure*}
%\center
\includegraphics[width=8cm]{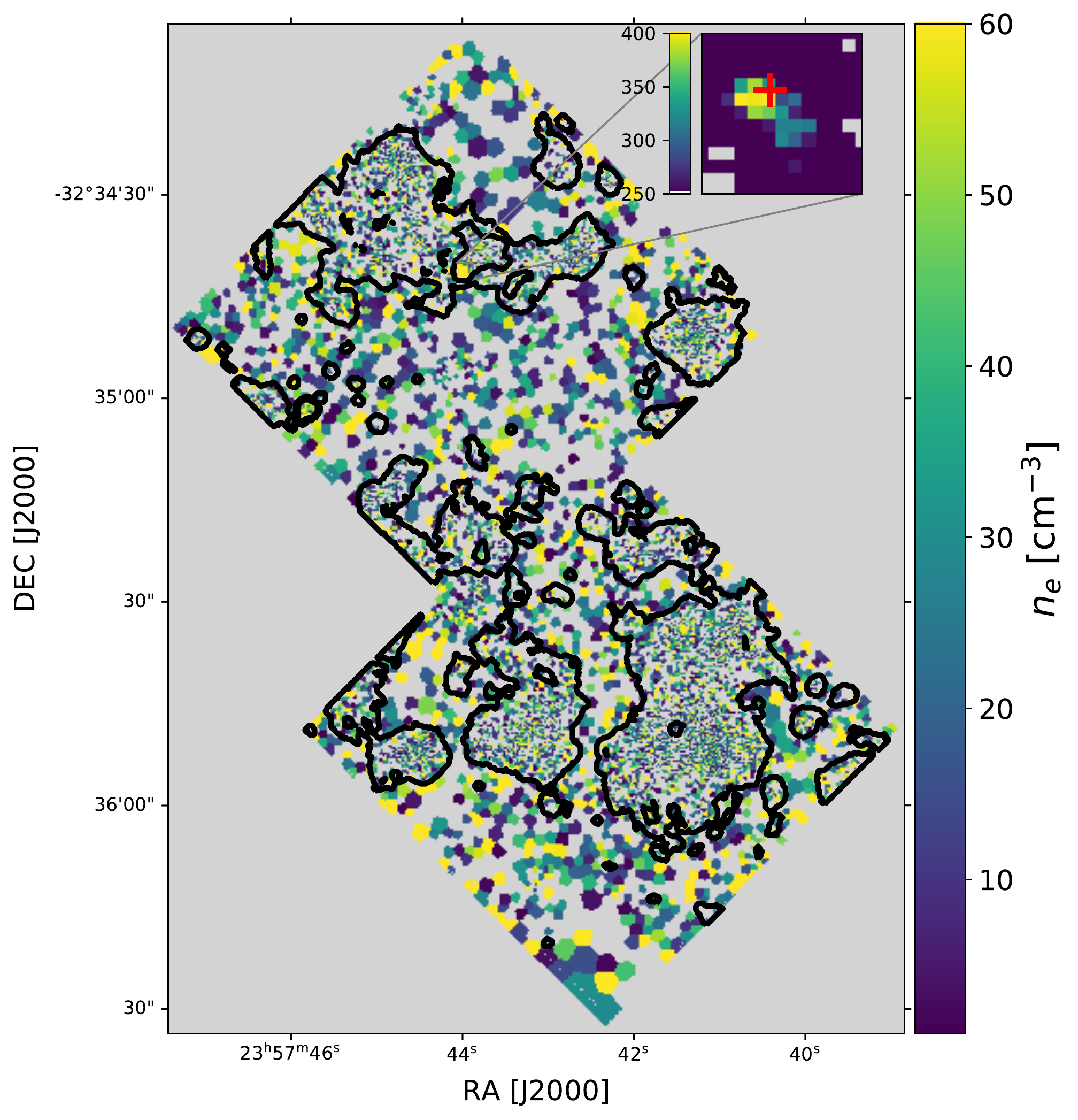}
\includegraphics[width=8cm]{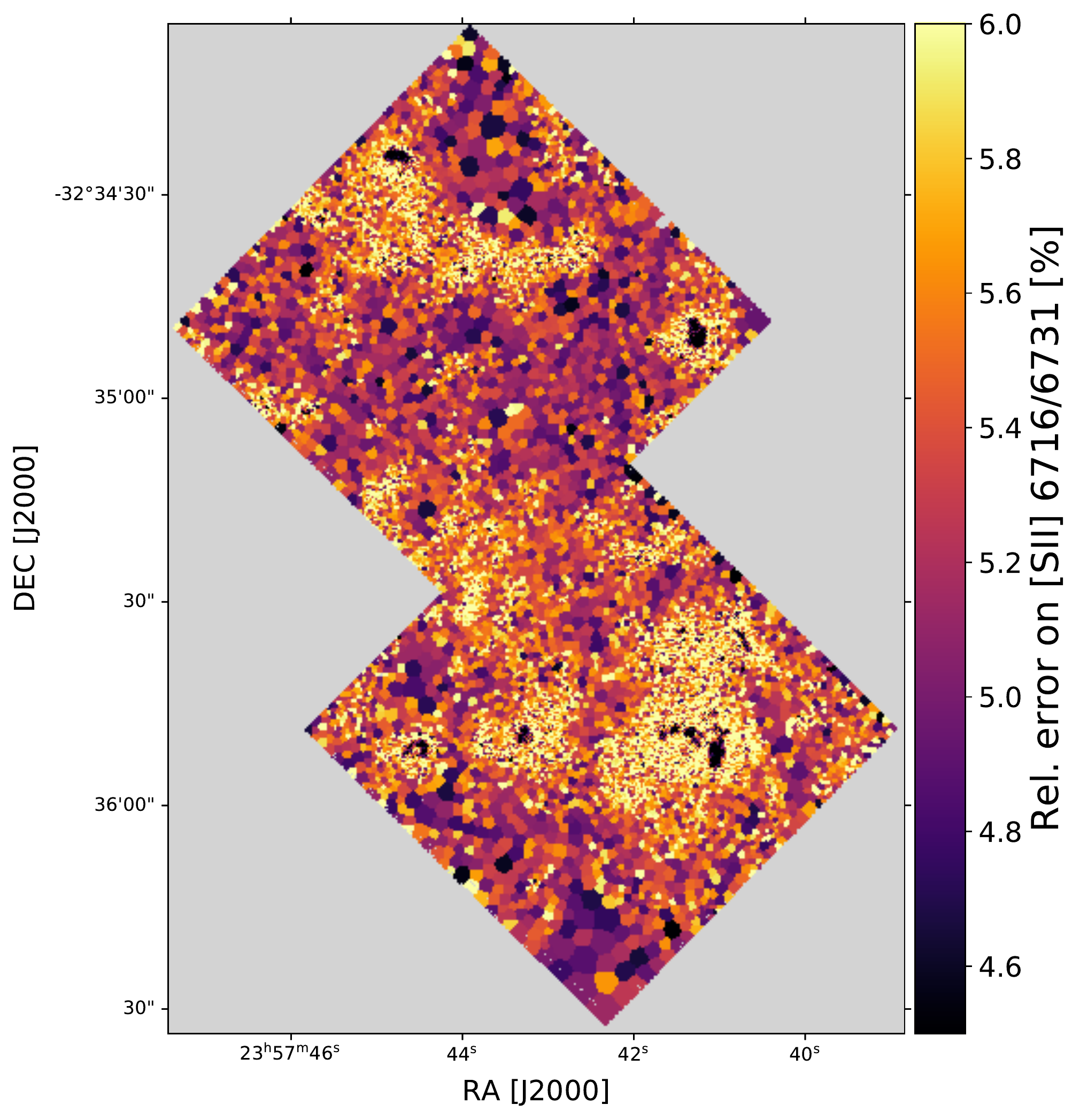}
\caption{Left: electron density computed with \textsc{pyneb}, for an assumed temperature $T_e \sim 8000$ K. HII region contours are overlaid in black. Grey-shaded areas indicate bins where the observed line ratio is below the sensitivity limit of the method (see bottom panel of Fig.~\ref{fig:emisgrid_s3}). The upper right panel shows a zoom-in into the planetary nebula candidate PN2 (red cross; see Table~\ref{table:candidates}). Right: relative error on the [SII]6716/6731 line ratio. The recovered values are below 7\% over all the FoV.}
\label{fig:ne}
\end{figure*}

\begin{figure}
\center
\includegraphics[width=9cm]{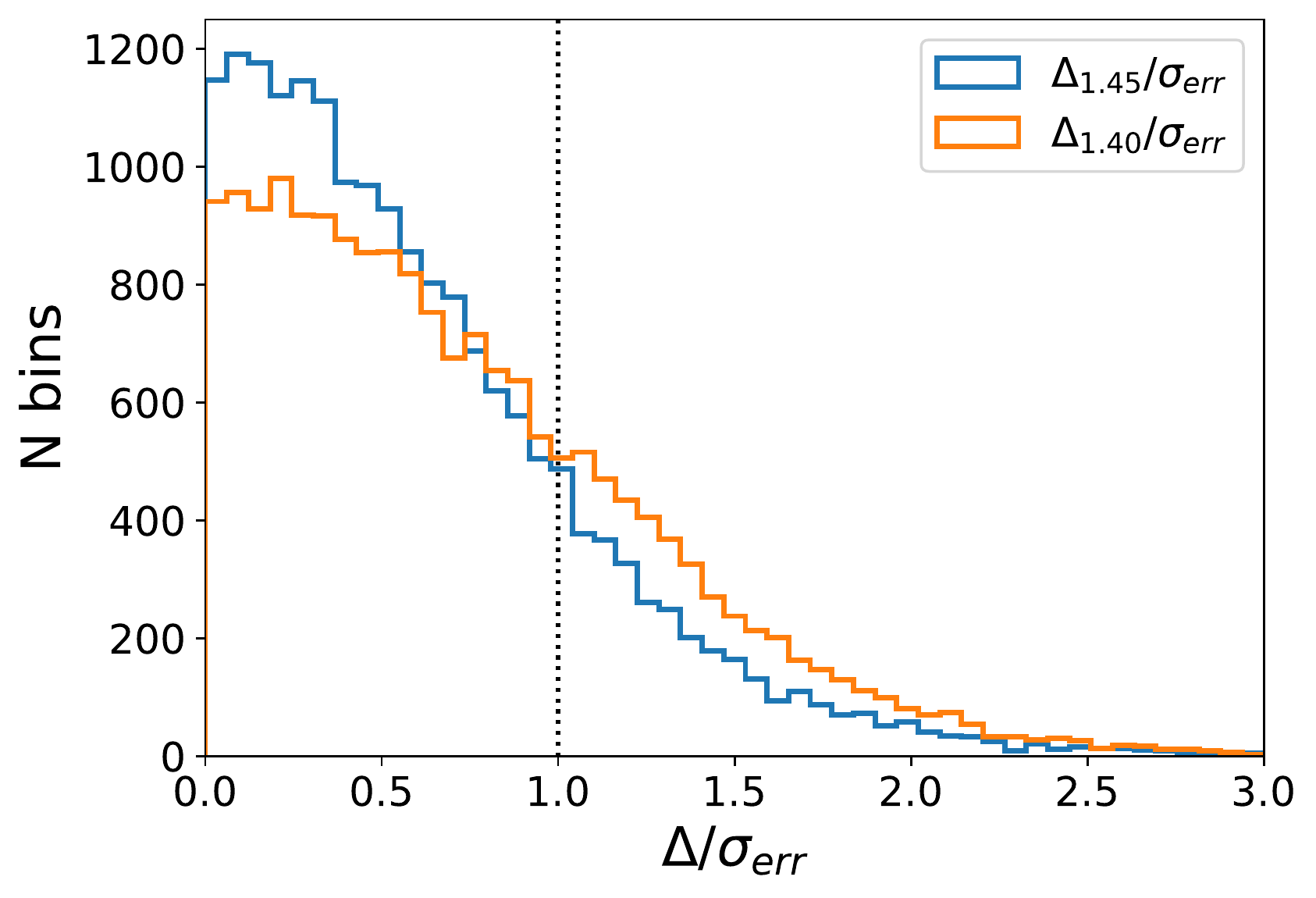}
\caption{Absolute distance of the observed [SII] ratio from a value of 1.45 ($\Delta_{1.45}$, blue histogram), corresponding to the low density limit $n_e = 1$ cm$^{-3}$, and from a value of 1.40 ($\Delta_{1.40}$, orange line), corresponding to a density $n_e \sim 30$ cm$^{-3}$, expressed in terms of the error on the line ratio $\sigma_{err}$.}
\label{fig:ne_hist}
\end{figure}

 We observe two density-sensitive lines in our data, [SII]6731 and [SII]6716. Before computing the line ratio, we bin the faintest line ([SII]6731) to cells with a S/N = 20. The same pattern is then applied to the other [SII] line. The recovered errors on the line ratio are below 7\% over all the FoV (right panel of Fig.~\ref{fig:ne}). We compute the density assuming a temperature $T_e \sim 8000$ K, corresponding to the median value observed across the FoV. The resulting density map is shown in Fig.~\ref{fig:ne} (left panel).
Densities range from 1 to 120 cm$^{-3}$, with an almost identical distribution for the DIG and HII regions. More than 50\% of bins within both HII regions and DIG remain with undetermined densities because the line ratio is below the sensitivity limit of the method. This is illustrated in the bottom panel of~Fig.~\ref{fig:emisgrid_s3}, where we show the relation between the line ratio and density as well as the range of observed values in the DIG (in green) and within the boundaries of the HII regions (in purple). The method is unable to recover densities below 1 cm$^{-3}$ (corresponding to a line ratio [SII]6716/6731 $\sim$ 1.45), and is largely insensitive in the range $n_e <  30$ ([SII]6716/6731 $\lesssim$ 1.40). Fig.~\ref{fig:ne_hist} shows indeed that, within 1$\sigma$ uncertainty on the line ratio, 81\% of the bins are consistent with $n_e \leq 1$ cm$^{-3}$, as well as 72\% are consistent with a density of $\sim 30$ cm$^{-3}$. This suggests that the gas densities within our FoV are largely consistent with densities below  $n_e \leq 30$ cm$^{-3}$.
Subtracting a median DIG spectrum from HII regions to account for diffuse emission on top of the regions does not impact our results.
The inset panel in the upper right corner of Fig.~\ref{fig:ne} shows a zoom-in into the planetary nebula candidate PN2 (see Table~\ref{table:candidates}). Densities at the location of the nebula and in the immediate surroundings are in the range 250-400 cm$^{-3}$, in agreement with the lower range probed by recent high-resolution studies of planetary nebulae in the Milky Way~\citep[see e.g.][]{walsh18, ibero19}.

\section{Properties and origin of the DIG}
\label{section:dig_properties}

In Sect.~\ref{section:analysis_gas}, we have classified the H$\alpha$ emitting gas into HII regions, bright regions ionised by other mechanisms (e.g by PNe and SNR) and DIG. Here we investigate the properties of the diffuse gas component.

\par
\begin{figure*}
\center
\includegraphics[width=16cm]{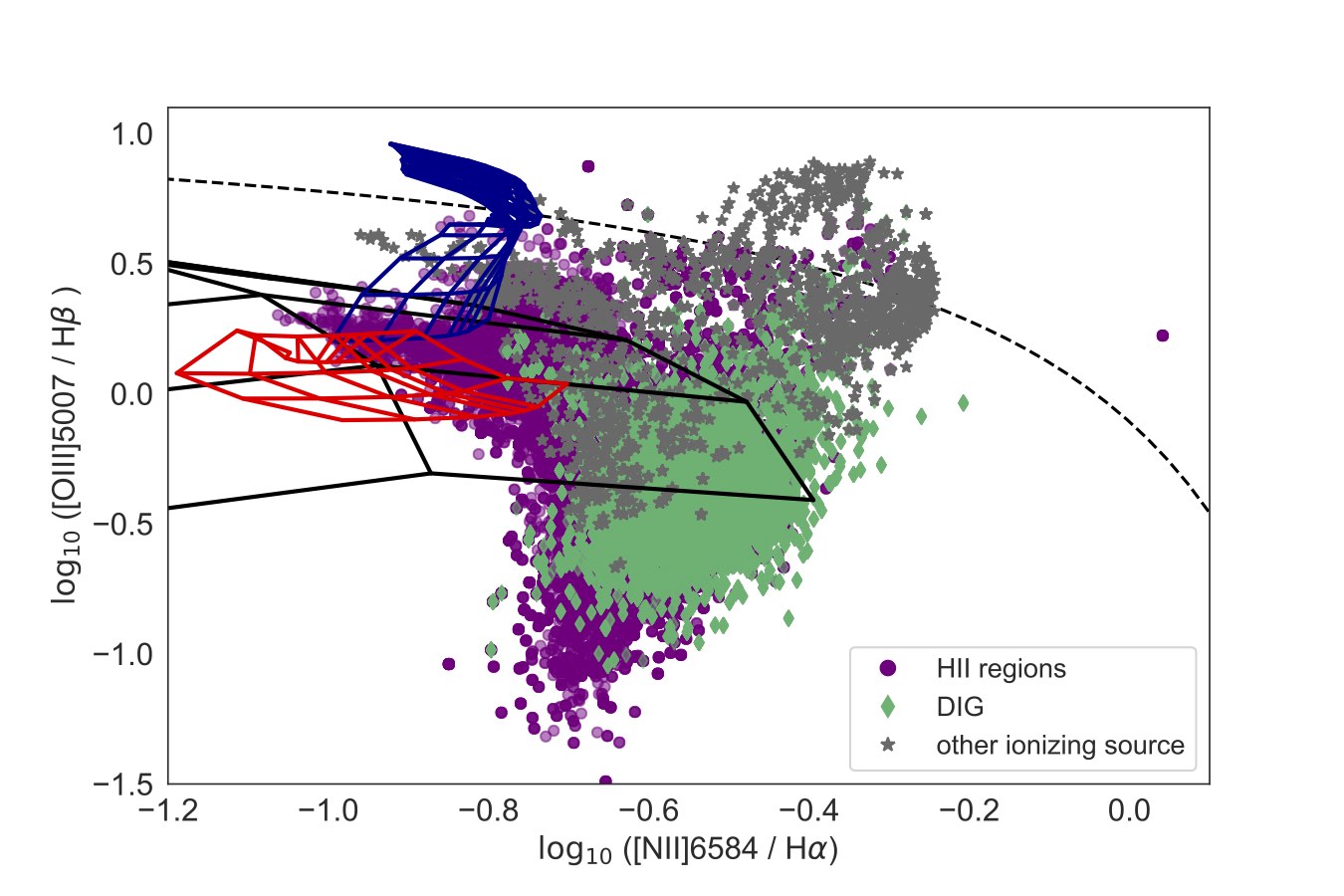}
\includegraphics[width=16cm]{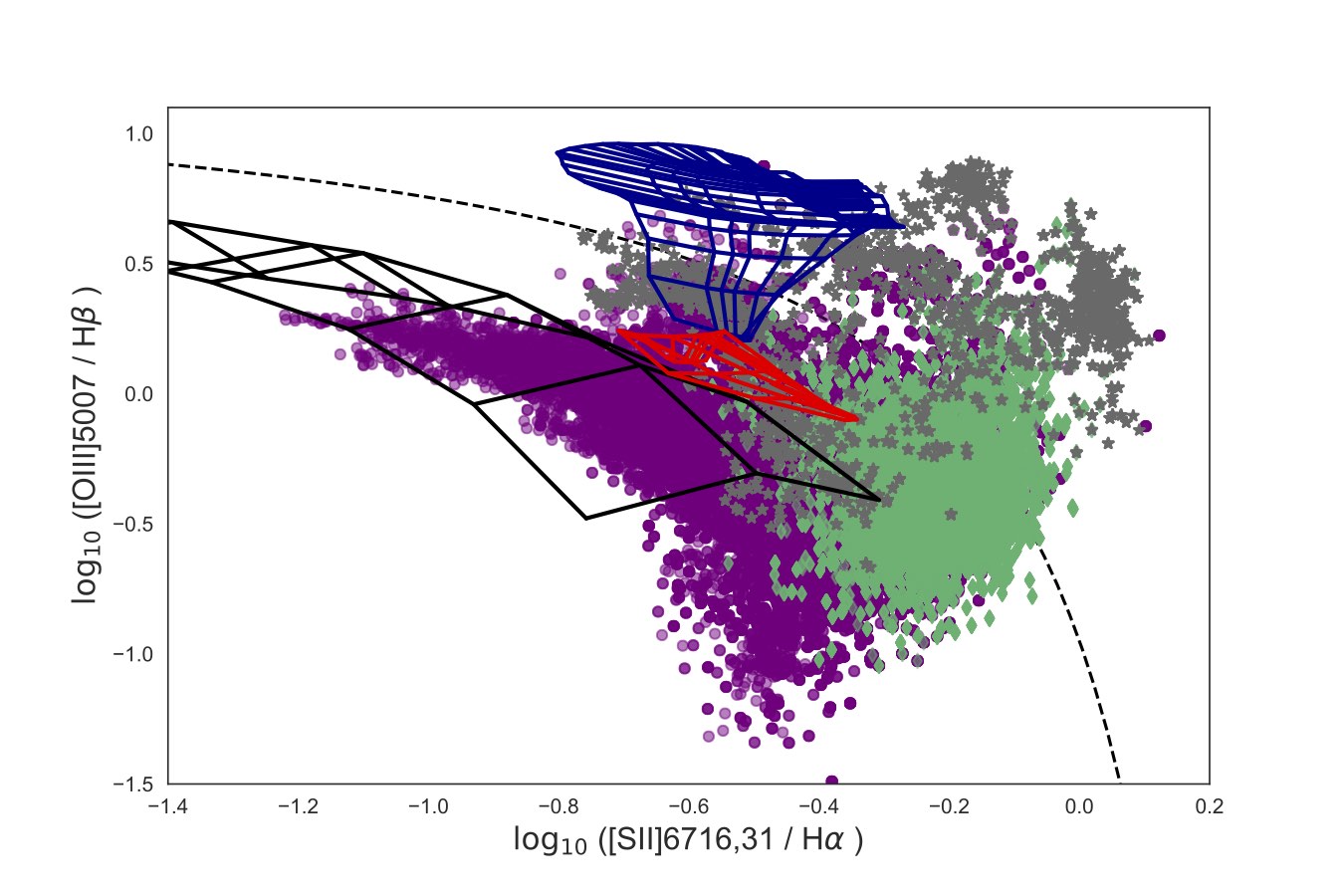}
\caption{[NII]- and [SII]-BPT diagrams with each spaxel colour-coded based on its spatial location: inside an HII region (purple circles), a region ionised by another source (grey stars) or in the DIG (green diamonds). Observations are compared with fast radiative shock models from \citet{allen08} (shock only models for $v_{shock} < 200$ in red; shock+precursor models for $v_{shock} > 200$ in dark blue) and models of star forming galaxies from~\citet{levesque10} (solid black). The black dashed line shows for reference the extreme starburst line from~\citet{kewley01}.}
\label{fig:bpt_all}
\end{figure*}

 By inverting our HII regions mask, we find a DIG fraction $f_{DIG} \sim$ 27\%  (42 \%) for regions selected on the H$\alpha$ (H$\alpha$/[SII]) map. Our two selection methods provide a lower and upper limit to the $f_{DIG}$. We study the ionised gas in resolved BPT diagrams. In the [NII]- and [SII]-BPT diagrams in Fig~\ref{fig:bpt_all}, each point corresponds to a spatial pixel (spaxel), and is colour-coded according to its belonging to an HII region, a region ionised by a different mechanism (see Sect.~\ref{section:outliers}) or the diffuse emitting gas. The data are compared to the models of low metallicity star forming galaxies from~\citet{levesque10} (see Sect.~\ref{section:hiiregions_selection}) and to fast radiative shock models from~\citet{allen08}. For the latter, we use the models with LMC metallicity ($Z \sim 0.008$) and $n_e = 1$, covering the range $b = (0.001 - 100)$ $\mu$G and $v_{shock}$ = 100 - 1000 km/s. Shock-only models are considered for  $v_{shock} <$ 200 km/s; otherwise the full shock+precursor models are used.
We observe that: firstly, the DIG exhibits enhanced [SII]/H$\alpha$ and [NII]/H$\alpha$ ratios, as observed in the Milky Way and other local galaxies \citep[see e.g.][]{haffner09}. Secondly, the bulk of the DIG is either consistent with the models from~\citet{levesque10} ([NII]-BPT) or just beyond the range spanned by the models, but below the extreme starburst line from~\citet{kewley01} ([SII]-BPT). Lastly, the fast shock models from~\citet{allen08} partially overlap with the regions featuring strong [OIII]5007/H$\alpha$ and [SII]6731/H$\alpha$ line ratios (colour-coded as `other ionising source' in Fig~\ref{fig:bpt_all}) as well as with some of the HII regions spaxels, but show very little overlap with the DIG regions. Using different assumptions for the metallicity and electron density in the shock models does not change this result.
%%%

\par
\begin{figure*}
\center
\includegraphics[width=18cm]{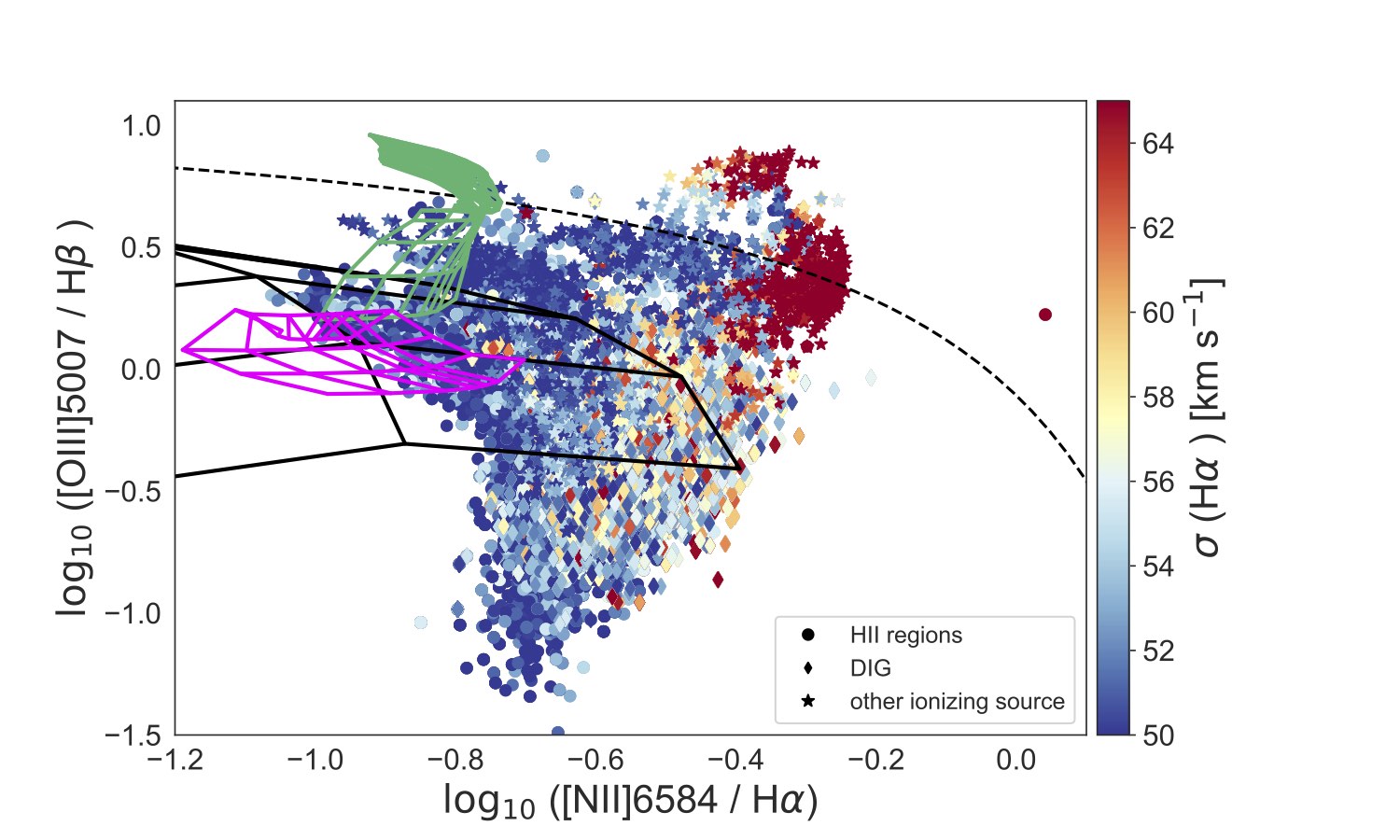} \\
\includegraphics[width=18cm]{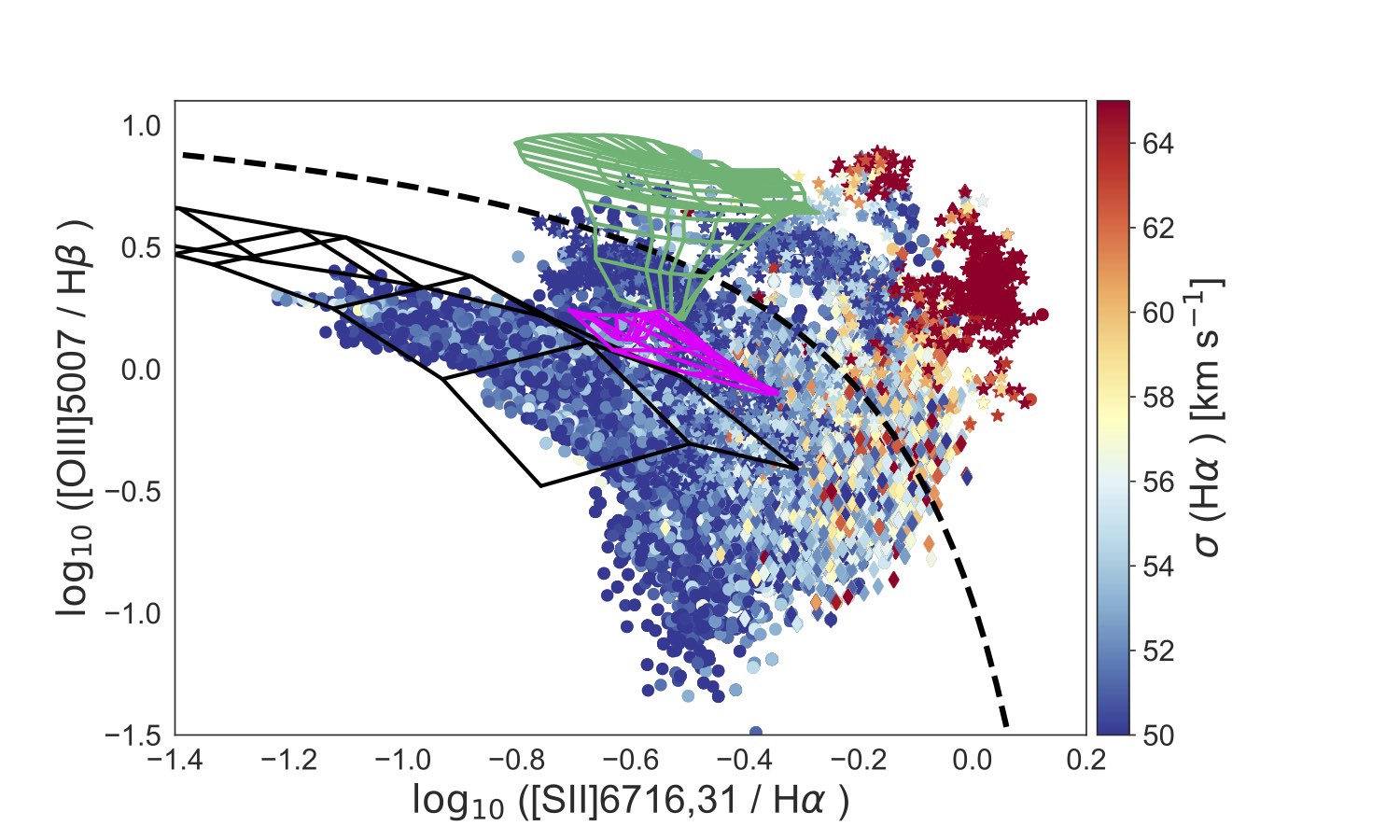}
\caption{As in Fig.~\ref{fig:bpt_all}, each point represents a spaxel; here the points are colour-coded by their H$\alpha$ velocity dispersion, whereas their shape varies depending on their location within HII regions (circles), DIG (diamonds) or strongly ionised regions (stars). The observed points are compared to fast radiative shock models from \citet{allen08} (shock only models for $v_{shock} < 200$ in magenta; shock+precursor models for $v_{shock} > 200$ in light green), models of star forming galaxies from~\citet{levesque10} (solid black) and the extreme starburst line from~\citet{kewley01} (dashed black).}
\label{fig:bptsigma}
\end{figure*}

\begin{figure}
\center
\includegraphics[width=11cm]{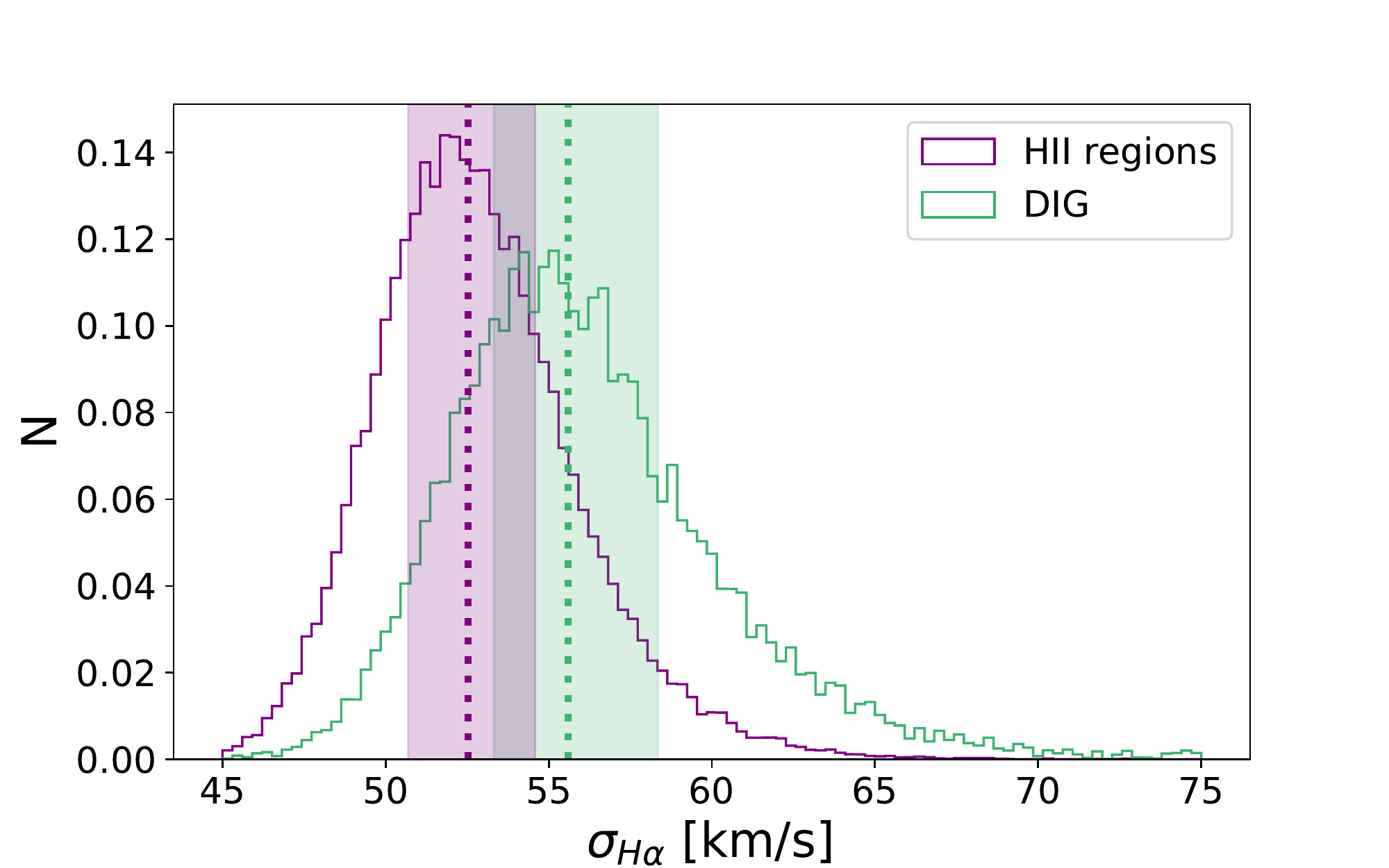}
\caption{Normalized distribution of the measured H$\alpha$ velocity dispersion for spaxels located in HII regions (purple) and in the DIG (green). The coloured area spans the first to third quartile of the distribution; the median is indicated as a dashed line.}
\label{fig:sigma_hist}
\end{figure}

 We also investigate the kinematics of the DIG.
In Fig.~\ref{fig:bptsigma} we plot a similar BPT diagram as in Fig.~\ref{fig:bpt_all}, but colour-code all spaxels by their H$\alpha$ velocity dispersion; the addition of the line of sight velocity dispersion as an ulterior parameter in the BPT diagram has been recently proposed by~\citet{oparin18}. In agreement with their work, we observe a clear trend between velocity dispersion and strength of the radiation field. In general, we observe that spaxels in other ionising sources (marked as stars in Fig.~\ref{fig:bptsigma}) exhibit the highest velocity dispersions, confirming that they are shock-dominated regions. Spaxels within HII regions have lower velocity dispersion than spaxels located outside. We also observe that spaxels labelled as DIG do overlap with HII region spaxels but have a larger range of velocity dispersion, as confirmed by Fig.~\ref{fig:sigma_hist}.

Finally, in Fig.~\ref{fig:sigma_halpha_dig}, we compare the location of the HII regions to the velocity dispersion of H$\alpha$, and in Fig~\ref{fig:sigma_hist} we show a histogram of the velocity dispersion measured in HII regions and DIG spaxels.
We recover
$\sigma_{H\alpha,HII} = 52.5_{-1.8}^{+2.1}$ \si{\kilo \metre \per \second} and $\sigma_{H\alpha,DIG} = 55.6_{-2.3}^{+2.7}$ \si{\kilo \metre \per \second}, where we have indicated the median and the first and third quartile of the distribution. A two-sample Kolmogorov–Smirnov (KS) statistical test recovers a p-value 
$\sim$ 0, indicating that the two distributions differ.
The tail of larger velocity dispersion in the DIG, which is not seen in HII regions, might be an indication of a higher turbulent motion in the ISM, and hinting to the fact that shocks and mechanisms other than photoionisation could still play a non-negligible role in the ionisation of the DIG. Alternatively, this could be due to a line of sight effect: DIG spaxels might have contributions from gas distributed along the line of sight, originating from regions with a physical separation up to a kpc, whereas HII regions spaxels are dominated by flux originating from a physically narrow region.

\section{Conclusions}
\label{section:conclusions}
We have presented recent MUSE AO observations of the nearby galaxy NGC 7793. We produced maps of nebular extinction, electron density and temperature, as well as stellar and gas kinematics. We analysed the state of the ionised gas and identified its various components: we constructed a sample of HII regions by applying a surface brightness cut in H$\alpha$ or in the line ratio H$\alpha$/[SII], and excluded contaminants by placing the resulting regions in BPT diagrams and inspecting their ratio of [OIII]5007 and [SII]6731 with respect to H$\alpha$. Using these two line ratios, we also identified two candidates PNe and three SNRs (two of which previously catalogued as candidates); all SNRs and one of the PNe also exhibit a large H$\alpha$ velocity dispersion. Finally, we produced a HeII 4686 linemap and identified nine WR stars candidates, seven of which were never catalogued before.

\par We then investigated the properties and origin of the DIG. Our main findings are as follows:
firstly, depending on the surface brightness cut applied in the selection of the HII regions, we obtain a DIG fraction $f_{DIG} \sim$ 27\% (H$\alpha$) or 42\% (H$\alpha$/[SII]). The diffuse gas exhibits enhanced [SII]/H$\alpha$ and [NII]/H$\alpha$ ratios, indicating its lower ionisation state, as broadly observed in nearby galaxies. Secondly, \textsc{pyneb} reveals that the line ratio [SII]6716/6731 in the majority of our FoV is consistent within 1$\sigma$ with electron densities below 30 cm$^{-3}$, with an almost identical distribution for the DIG and HII regions. The electron temperature of the DIG is $\sim$ 3000 K higher than in HII regions, in agreement with DIG studies in the Milky Way. We moreover observe an apparent inverse dependence of the electron temperature on the H$\alpha$ surface brightness, consistent with what observed in the Milky Way, where the temperature of the DIG increases for decreasing emission measure~\citep{haffner09}. However, due to the sensitivity limit in the weak [SIII]6312 line, we cannot rule out the existence of a population of lower H$\alpha$ surface brightness regions at low $T_e$.
Thirdly, by comparing the DIG emission with models for low metallicity star forming galaxies from~\citet{levesque10}, we observe that the bulk of the DIG emission is consistent with being photoionised or just above the range spanned by the models. We also assess that the fast radiative shock models from~\citet{allen08} show little to no overlap with the DIG spaxels.
Finally, we add as third variable in our BPT diagram the line of sight velocity dispersion of the H$\alpha$ line, as recently proposed by~\citet{oparin18}, and observe a clear trend between the velocity dispersion and the strength of the radiation field. In particular, we find enhanced gas velocity dispersions in the DIG with respect to HII regions, indicating that effects others than photoionisation might play a non-negligible role.

In Paper II, we will combine these results on the ionised gas with the stellar population studied within the HST--LEGUS programme and the dense gas properties provided by the ALMA-LEGUS data. We will further investigate how the physical properties of HII regions (luminosity, geometry, density or ionisation boundness) relate to the ages and masses of the hosted stellar clusters and the number of massive stars and WR candidates. We will also provide a quantitative estimate of which fraction of the DIG luminosity could be accounted for by ionising radiation escaping HII regions.

\begin{acknowledgements} The authors thank the Anonymous Referee for the helpful comments and constructive remarks on this manuscript.
This work is based on observations collected at the European Southern Observatory under ESO programme 60.A-9188(A).
A.A. acknowledges the support of the Swedish Research Council, Vetenskapsr\aa{}det, and the Swedish National Space Agency (SNSA). GB acknowledges financial support from DGAPA-UNAM through PAPIIT project IG100319. MF acknowledges support by the Science and Technology Facilities Council [grant number  ST/P000541/1]. This project has received funding from the European Research Council (ERC) under the European Union's Horizon 2020 research and innovation programme (grant agreement No 757535). AW acknowledges financial support from DGAPA-UNAM
through PAPIIT project  IA105018.  This research made use of Astropy\footnote{http://www.astropy.org}, a community-developed core Python package for Astronomy \citep{astropy:2013, astropy:2018}. PyRAF is a product of the Space Telescope Science Institute, which is operated by AURA for NASA. \end{acknowledgements}

%\nocite{*}
\bibpunct{(}{)}{;}{a}{}{,} % to follow the A&A style
\bibliographystyle{aa}
\bibliography{dellabruna_2020}
\afterpage{\clearpage}

\begin{appendix}
\section{Spectra of PNe, SNR and WR candidates}
\label{section:appendix}

\begin{figure*}
\includegraphics[width=18cm]{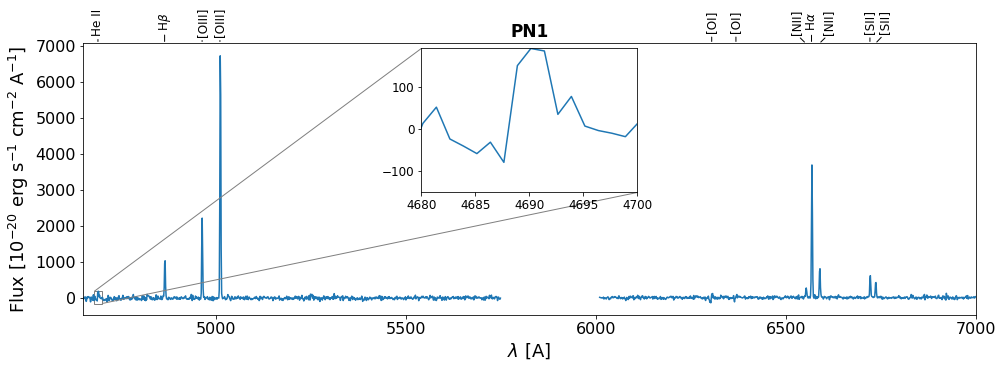}
\includegraphics[width=18cm]{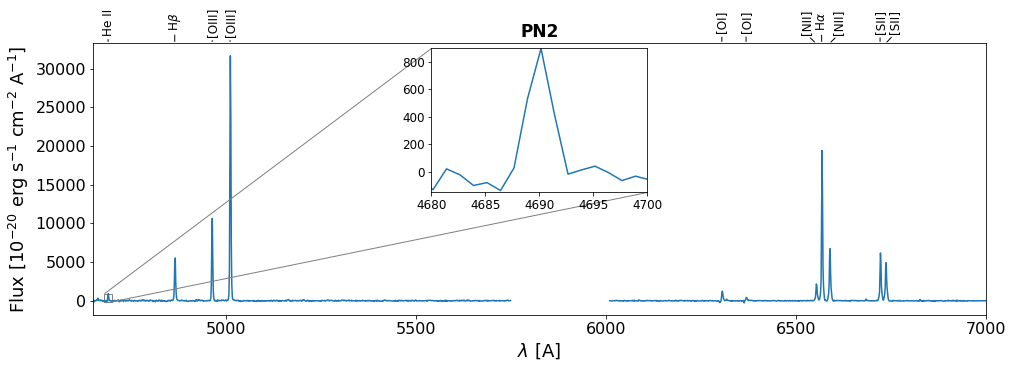}
\caption{Spectra of the candidate PNe listed in Table \ref{table:candidates}, extracted from the gas cube using a circular aperture of 0.4\arcsec. The spectra feature HeII 4686 emission as well as a very strong [OIII]5007 emission relative to H$\alpha$.}
\label{fig:pn_spectra}
\end{figure*}

\begin{figure*}
\includegraphics[width=18cm]{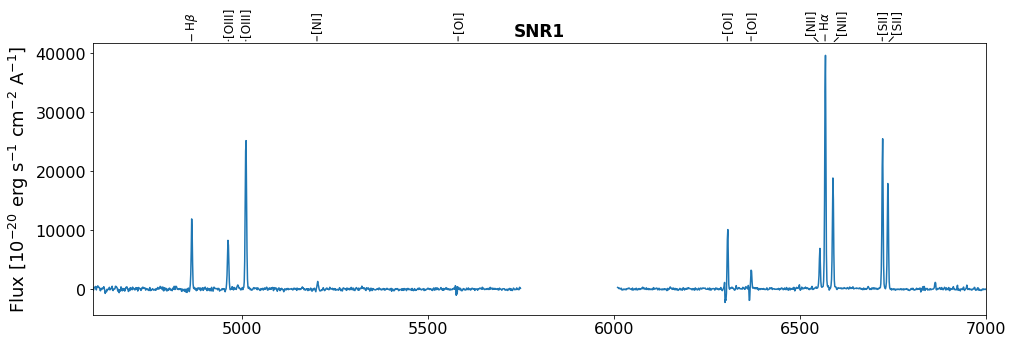}
\includegraphics[width=18cm]{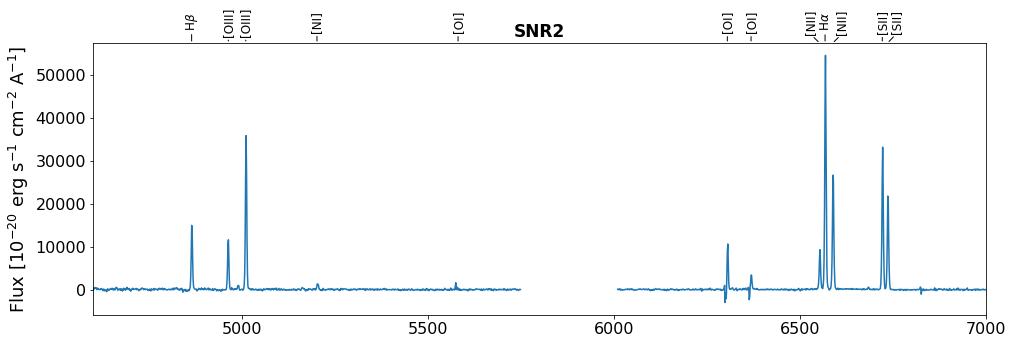}
\includegraphics[width=18cm]{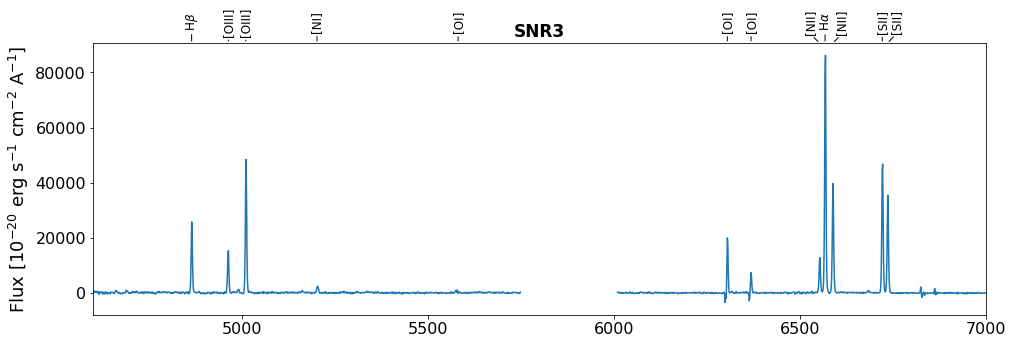}
\caption{Spectra of the candidate SNR listed in Table \ref{table:candidates}, extracted from the gas cube using a circular aperture of 1.4\arcsec. All candidates feature an enhanced [SII]6731/H$\alpha$ ratio as well as relatively strong [OI]6300,6364 emission (both contaminated by residual sky emission). We also observe a weak [NI]5198 emission.}
\label{fig:snr_spectra}
\end{figure*}

\begin{figure*}
\includegraphics[width=18cm]{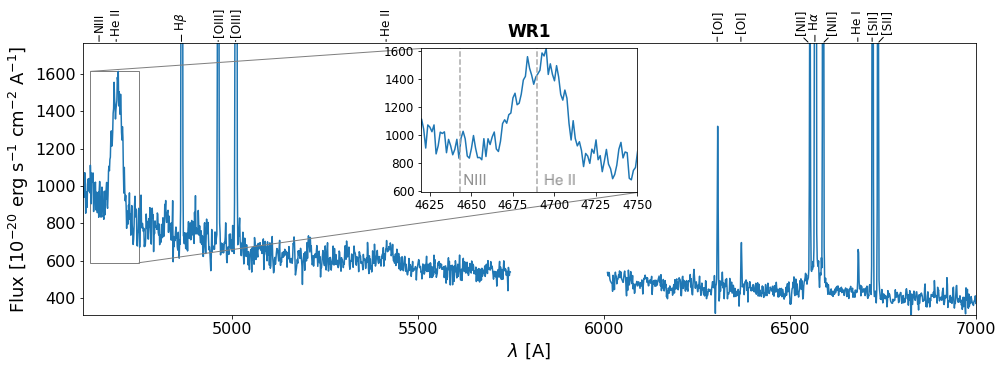}
\includegraphics[width=18cm]{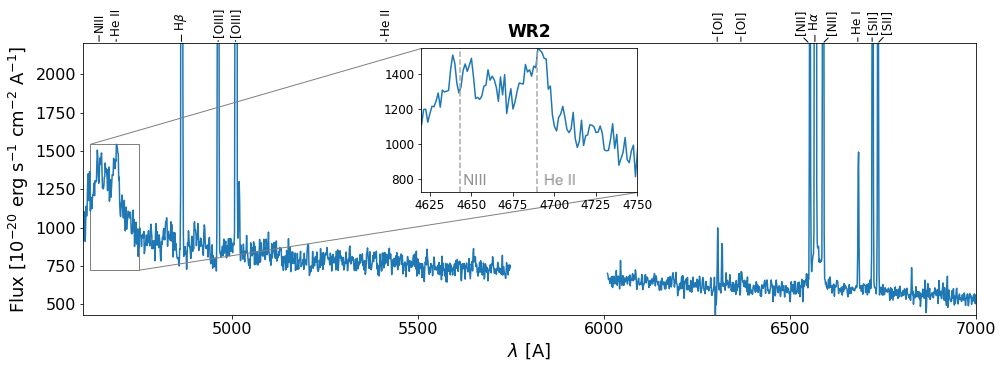}
\includegraphics[width=18cm]{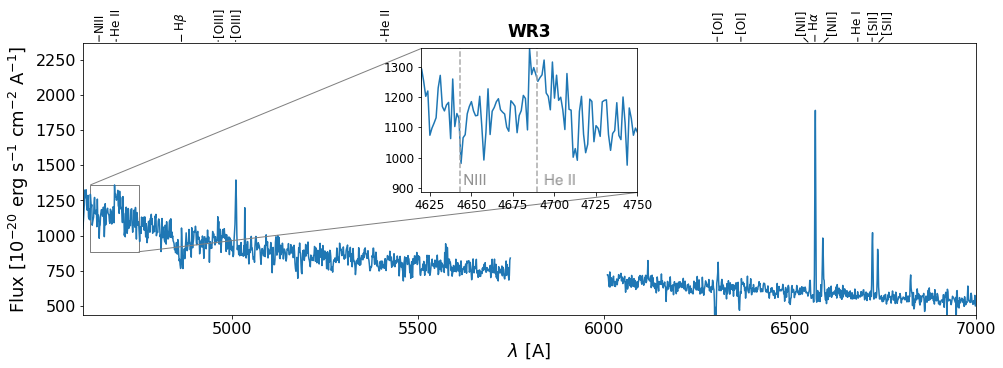}

\caption{Spectra of the candidate WR listed in Table \ref{table:candidates}, extracted from the full cube using a circular aperture of 0.4\arcsec. One can observe the characteristic HeII 4686 `bump' as well as HeII 5411, [OI]6300,6364 (both contaminated by residual sky) and HeI 6678 emission. Some of the candidates also feature NIII 4640 emission.}
\label{fig:wr_spectra}
\end{figure*}

\begin{figure*}
\includegraphics[width=18cm]{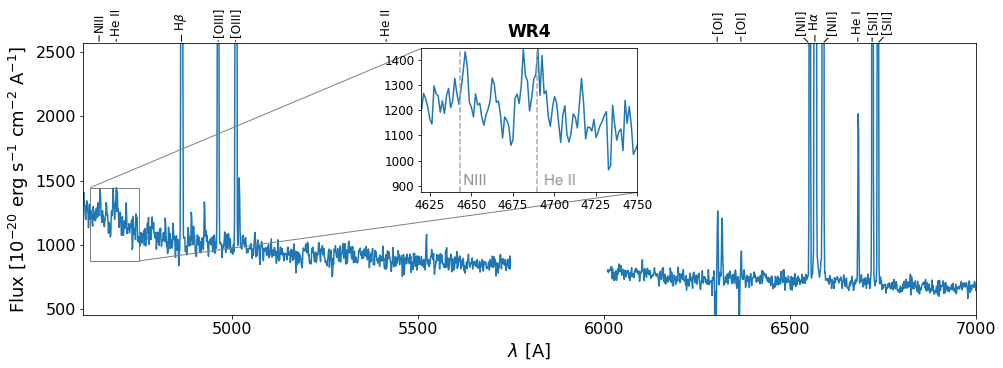}
\includegraphics[width=18cm]{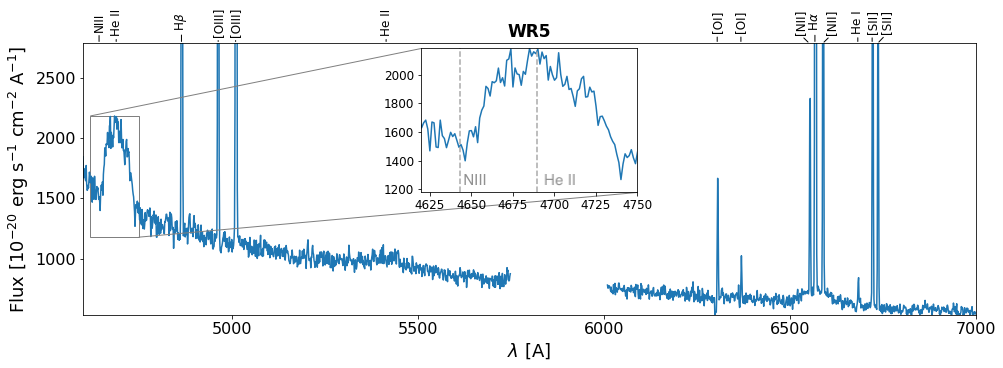}
\includegraphics[width=18cm]{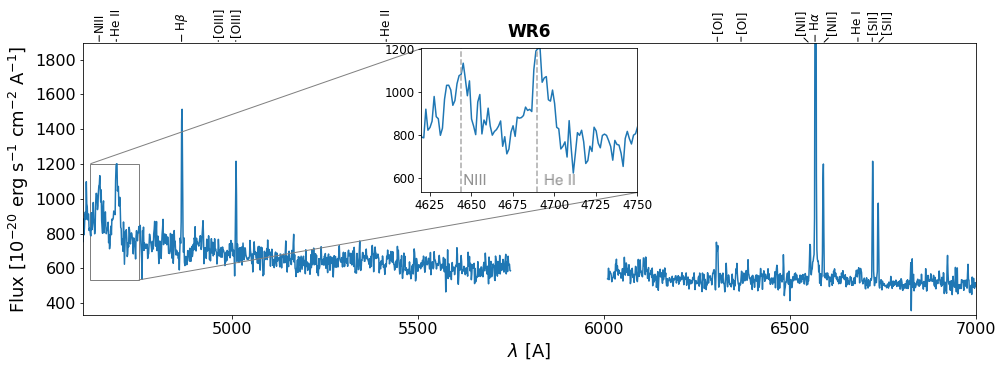}

\caption{Fig.~\ref{fig:wr_spectra} continued.}
\end{figure*}

\begin{figure*}
\includegraphics[width=18cm]{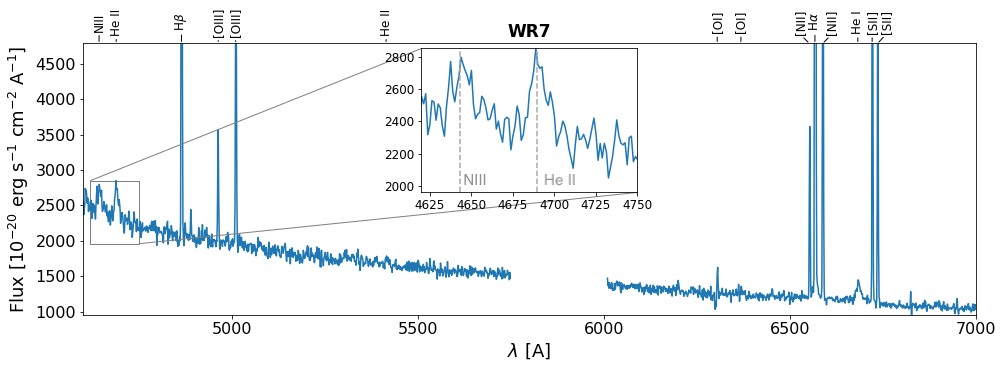}
\includegraphics[width=18cm]{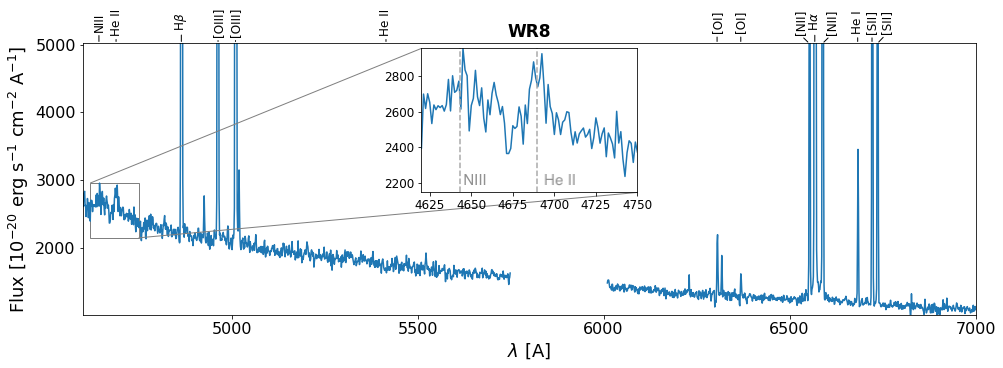}
\includegraphics[width=18cm]{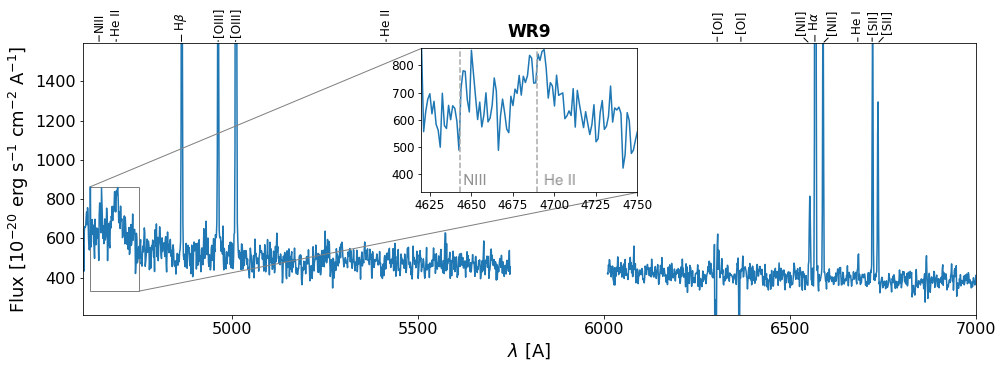}

\caption{Fig.~\ref{fig:wr_spectra} continued.}
\end{figure*}

\end{appendix}
\end{document}